\documentclass[aps,prb,twocolumn,floatfix]{revtex4-2}

\usepackage{amssymb}
\usepackage{amsmath}
\usepackage{graphicx}

\begin{document}

\title{Critical current in thin flat superconductors
 with Bean-Livingston and geometrical barriers}

\author{G.~P.~Mikitik}
\affiliation{B.~Verkin Institute for Low Temperature Physics \&
Engineering, Ukrainian Academy of Sciences, Kharkov 61103,
Ukraine}

\begin{abstract}
Dependence of the critical current $I_c$ on the applied magnetic field $H_a$ is theoretically studied for a thin superconducting strip of a rectangular cross section, taking an interplay between the Bean-Livingston and the geometric barriers in the sample into account. It is assumed that bulk vortex pinning is negligible, and the London penetration depth $\lambda$ is essentially less than the thickness $d$ of the strip. To investigate the effect of these barriers on $I_c$ rigorously, a two-dimensional distribution of the current over the cross section of the sample is derived, using the approach based on the methods of conformal mappings. With this distribution, the dependence $I_c(H_a)$ is calculated
for the fields $H_a$ not exceeding the lower critical field.
This calculation reveals that the following two situations are possible: i) The critical current $I_c(H_a)$ is determined by the Bean-Livingston barrier in the corners of the strip. ii) The geometrical barrier prevails at low $H_a$, but with increasing magnetic field, the Bean-Livingston barrier begins to dominate. The realization of one or the other of these two situations is determined by the ratio $\lambda/d$.
\end{abstract}

\maketitle

\section{Introduction}

The Bean-Livingston \cite{BL} and geometrical \cite{ZL} barriers
are important for understanding many phenomena in type-II
superconductors. In particular, these barriers lead to a
hysteretic magnetic behavior of the superconductors even in
absence of any bulk pinning of vortices
\cite{BL,ZL,ZL1,ternov,clem,ben1,br1,br2,willa}. They also influence the magnetic relaxation \cite{bur1,bur2} and transport properties of the superconductors \cite{likh,bur3,ben2,mak-c}.
Various manifestations of the Bean-Livingston and geometrical
barriers were experimentally studied in numerous works
\cite{Kop90,Kon91,Chik92,ind,Mar95,Maj95,Kim95,Fli95,Mor96,Mor97,
Pal98,Fuchs98,Fuchs98b,Fuchs98a,Mish00,geim,plourde,Xiao02,Olsen,Lya04,Haim09,johns,
Segev11,shen,prib,embon,dolz,dobrov,kuro}.
However, it was demonstrated in Ref.~\cite{jetp13} that an
{\em interplay} between these barriers should have a pronounced effect on any phenomenon associated with the vortex penetration into a superconductor. Below  we theoretically study how this interplay
influences the dependence of $I_c$, the critical current of a platelet-shaped type-II superconductor, on the applied magnetic field $H_a$  perpendicular to the plane of the sample. For  simplicity, we assume that flux-line pinning is negligible  in the superconductor.

\begin{figure}[t] 
 \centering  \vspace{+9 pt}
\includegraphics[bburx=720,bbury=270,scale=.45]{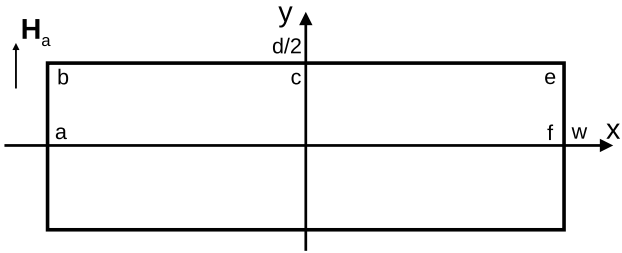}
\caption{\label{fig1} The rectangular cross section of the infinitely long superconducting strip of the width $2w$ and of the thickness $d$. The magnetic field $H_a$ is applied along the $y$ axis, and the current $I$ flows in the positive direction of the $z$ axis (in the direction $\hat{\bf z}=[\hat{\bf x} \times \hat{\bf y}]$ where $\hat{\bf x}$, $\hat{\bf y}$ are the unit vectors along the $x$ and $y$ axes, respectively). The points a, b, c, e, f on the surfaces of the strip correspond to the following values of the parameter $u$: $-1/\sqrt{1-m}$, $-1$, $0$, $1$, and $1/\sqrt{1-m}$, respectively.
 } \end{figure}

As is known, the Bean-Livingston barrier in bulk superconductors is due to the attraction of a penetrating vortex to the sample surface at the distances of the order of the London penetration depth $\lambda$ \cite{BL}. The geometrical barrier has another origin, and it is due to the shape of the superconductor \cite{ZL,ind}. This barrier appears only for the samples different from an ellipsoid.
In particular, in the platelet-shaped superconductors the position-dependent energy of a  penetrating vortex sharply increases near the edges due to the increase of the vortex length from zero to the sample thickness $d$ and decreases toward the center of the platelet due to the work of the Meissner currents.
It is necessary to emphasize that the interplay of these barriers occurs only in the bulk superconductors when $\lambda \ll d$.
In the case of thin superconducting films for which their
thickness $d$ is essentially less than the London penetration
depth $\lambda$, the attraction of a vortex to the film edges
develops on the scale noticeably larger than the effective
penetration depth $\lambda_{\rm eff}=\lambda^2/d \gg \lambda, d$
\cite{kog} whereas the effect of the vortex-length variation (i.e., the geometrical barrier) is not essential in this case. Hence, only one type of the barrier exists in this situation, and it was named as the extended Bean-Livingston barrier \cite{jetp13}. In this paper we shall study the case $\lambda\ll d$ only. For simplicity, we consider a thin superconducting strip of a rectangular cross section of width $2w$ ($-w\le x\le w$) and thickness $d$ ($-d/2 \le y \le d/2$; $d\ll w$) which infinitely extends in the $z$ direction (Fig.~\ref{fig1}). The strip is subjected to a perpendicular applied magnetic field ${\bf H}_a=(0,H_a,0)$, and it carries a total current $I$ in the positive $z$ direction.

In the familiar  approach to the calculation of the critical current  $I_c(H_a)$ \cite{likh,ben2,mak-c}, the sample is considered as an infinitely thin strip, and the barriers are modeled by the condition that the Lorentz force near the appropriate edge of the strip should reach a certain critical value for vortices to penetrate into the sample. At $d\gg \lambda$, if only the geometrical barrier is considered, this approach leads to the following estimate of $I_c$ at $H_a=0$ \cite{ben2}:
 \[
 I_c(0)\approx  \pi\sqrt{2wd}\,H_{c1},
 \]
where $H_{c1}$ is the lower critical field. Within this approach, the existence of the Bean-livingston barrier is taken into account by the replacement of $H_{c1}$ by some phenomenological field $H_b$ lying in the interval $H_{c1}\le H_b \le H_c$ \cite{ben2} where $H_c=H_{c1} \sqrt{2} \kappa /\ln\kappa$ is the thermodynamic critical field, and $\kappa$ is the Ginzburg-Landau parameter.  On the other hand in the case $d\ll \lambda$, one has:
 \[
 I_c(0)\approx \pi C\sqrt{2w\lambda_{\rm eff}}\, j_{dp}d\sim \pi\sqrt{2wd}\,H_{c},
 \]
where the numerical factor $C$  lies in the interval from $\sqrt{2/\pi}$ to $\sqrt{2}$  \cite{likh,mak-c,plourde,LO,aslam,bezug}, and $j_{dp}= (2/3)^{3/2}H_c/\lambda$ is depairing current density \cite{bl}. It is seen that the expressions for $I_c(0)$ derived in the regions $d\gg \lambda$ and $d\ll \lambda$ can agree at the boundary of these regions, $d\sim \lambda$, only if the relative role of the Bean-Livingston barrier increases with decreasing $d$, and  $H_b$ reaches $H_c$ at this boundary. However, in order to investigate if the decrease in $d$ really enhances the role of the Bean-Livingston barrier and how this interplay of the barriers influences the critical current $I_c$, one cannot neglect the thickness of the strip near its edges {\it even in the case of thin samples}. In this paper a two-dimensional distribution of the current over the cross section of the strip is found that permits one to answer these questions. In obtaining the distribution, the approach of Refs.~\cite{Meissner,jetp13} is exploited which is based on the methods of conformal mappings. For simplicity, we shall imply
below that the superconductor is isotropic and shall restrict our consideration to the region of the applied magnetic fields, $0\le H_a\lesssim H_{c1}$, in which the effect of the barriers on the critical current is most pronounced.

The paper is structured as follows: In Sec.~\ref{II} we present
the two dimensional distributions of the currents in the
strip with the rectangular cross section. The strip is either in the Meissner state or in the state with a vortex dome. The results of this section are valid not only for the thin strips but also for samples with an arbitrary aspect ratio $d/2w$. Using the distributions of the currents, in Sec.~\ref{III} we analyze the Bean-livingston and geometrical barriers in the thin strips, and derive conditions of vortex entry into the sample and vortex exit from it. Two scenarios of the vortex entry are also discussed there. Using the vortex entry and exit conditions, the critical current $I_c(H_a)$ of the strips is calculated in Sec.~\ref{IV}. In Sec.~\ref{V} we discuss the results of experiments \cite{Pal98,Fuchs98,Fuchs98b,Haim09} and possibility to detect an  unusual vortex state in the strip. The obtained results are briefly summarized in Conclusions, and the Appendices contain some mathematical details of the calculations.

\section{Surface currents in the strip with rectangular cross section} \label{II}

\subsection{Strip in the Meissner state}

For the strip in the Meissner state, the magnetic field ${\bf
H}(x,y)$ outside the sample can be found from the Maxwell
equations ${\rm div}{\bf H}=0$ and ${\rm rot}{\bf H}=0$, and hence
the field can be described both by the scalar potential
$\varphi(x,y)$, ${\bf H}=-\nabla \varphi$, and by the vector
potential ${\bf A}=\hat{\bf z} A(x,y)$, ${\bf H}={\rm rot}{\bf A}$,
where $\hat{\bf z}$ is the unit vector along the z axis. The complex
potential $\varphi-i A$ is known to be an analytical
function of $x+iy$ \cite{LL}. For the strip with the rectangular cross section and with nonzero $H_a$ and $I$, this potential was obtained with a conformal mapping \cite{Meissner}. Calculating ${\bf H}=-\nabla \varphi$ with the use of the obtained potential at the surface of the strip (${\bf H}$ is tangential to the surface in the Meissner state), one finds the Meissner sheet currents $J_M=J_z$ flowing near this surface in the layer of the thickness  $\sim \lambda$,
\begin{eqnarray} \label{1}
{\bf J}_M=[{\bf n}\times {\bf H}],
\end{eqnarray}
where ${\bf n}$ is the outward normal to the surface of the sample  at the point of interest \cite{LL}.

\begin{figure}[t]
 \centering  \vspace{+9 pt}
\includegraphics[scale=.48]{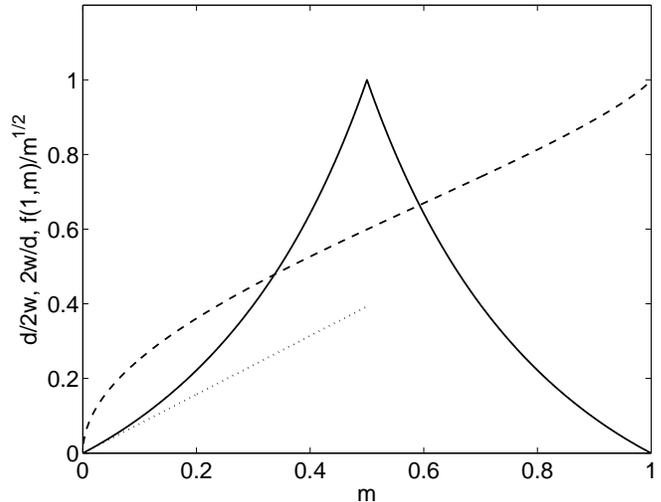}
\caption{\label{fig2} The dependences of $d/2w$ on $m$ at $0<m\le 1/2$ and of $2w/d$ on $m$ at $1/2 \le m <1$ according to formula (\ref{4}) (the solid line), the dotted line depicts the function $d/2w=\pi m/4$ which corresponds to Eq.~({5}). The dashed line shows the $m$-dependence of the ratio $f(1,m)/\sqrt{m}$ that appears in Eq.~(\ref{9}).
   } \end{figure}

\begin{figure}[b] 
 \centering  \vspace{+9 pt}
\includegraphics[scale=.48]{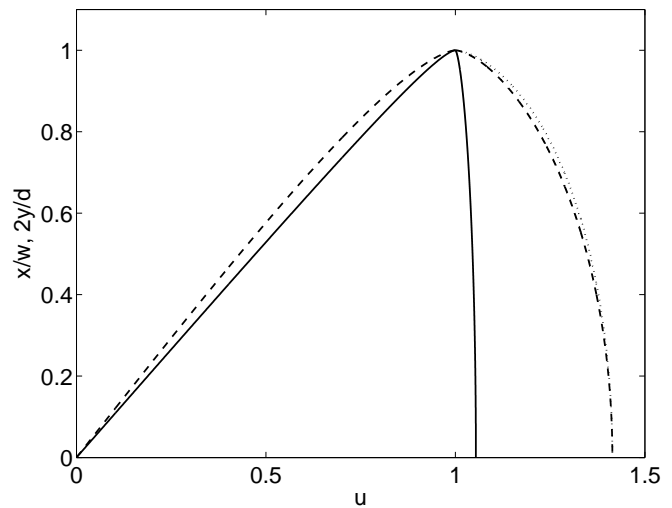}
\caption{\label{fig3} The dependences of $x/w$ and $2y/d$ on the parameter $u$ at $m=0.1$ (the solid line) and $m=0.5$ (the dashed line) according to formulas (\ref{2}), (\ref{3}), (\ref{6}), (\ref{7}). The parameter $u$ runs from $0$ to $1/\sqrt{1-m}$. The dotted line  corresponds to Eqs.~(\ref{7}) and (\ref{8}) at $m=1/2$.
   } \end{figure}

In the case $H_a \neq 0$, $I=0$, the above-mentioned mapping was detailed in Ref.~\cite{jetp13} and is presented in the Supplemental Material \cite{SuppMat}, whereas in the case $H_a=0$, $I \neq 0$, the mapping of the exterior of a circle to the exterior of a rectangle in the $x+iy$ plane reduces the problem for the strip to that of a cylindrical wire. In the general case, $H_a \neq 0$ and $I\neq 0$, the current $J_M$ is a superposition of the currents in these two specific cases. To represents the obtained results, it is convenient to parameterize the surface of the strip by a single variable. Since at $I=0$, $H_a\neq 0$ the Meissner currents are symmetric about the $x$ axis and antisymmetric about the $y$ axis, it is sufficient to deal with a quarter of the surface of the strip, (e.g., $x \ge 0$, $y\le 0$) and to parameterize it with the single variable $t$ changing from $0$ to $1/\sqrt m$ \cite{jetp13}. Here $m$ is a constant parameter, $0\le m \le 1$, the value of which is determined by the aspect ratio of the strip, $d/2w$, see below. However, in the general case when both $I\neq 0$ and $H_a\neq 0$, only the reflection symmetry of the currents about the $x$ axis  persists, and so in this paper we parameterize  the upper half of the surface of the strip ($y\ge 0$) by the single variable $u$. This $u$ changes from $-1/\sqrt{1-m}$ to $1/\sqrt{1- m}$ \cite{com1} (Fig.~\ref{fig1}). In particular, the upper surface of the strip ($-w\le x\le w$, $y=d/2$) is  parameterized as follows ($-1\le u \le 1$):
\begin{eqnarray} \label{2}
\frac{x}{w}=\frac{f(u,1-m)}{f(1,1-m)}
\end{eqnarray}
where
\begin{eqnarray} \label{3}
f(u,m)&\equiv& m\int_0^{u}\!\!\frac{\sqrt{1-v^2}}{\sqrt{1-mv^2}}\,dv \nonumber\\ &=&E(\varphi,k)-(k')^2F(\varphi,k),
\end{eqnarray}
$k=\sqrt{m}$, $k'=\sqrt{1-m}$, $\varphi=\arcsin(u)$, $F(\varphi,k)$ and $E(\varphi,k)$ are the incomplete elliptic integrals of the first and second kinds, respectively.
The points $u=\pm 1$ correspond to the upper corners of the strip, $(\pm w,d/2)$. The constant parameter $m$ is found from the equation:
\begin{eqnarray} \label{4}
\frac{d}{2w}=\frac{f(1,m)}{f(1,1-m)}=\frac{E(k)- (k')^2K(k)}{E(k')- k^2 K(k')},
\end{eqnarray}
where $K(k)\equiv F(\pi/2,k)$ and $E(k)\equiv E(\pi/2,k)$ are the complete elliptic integrals. The solution of this equation is presented in Fig.~\ref{fig2}. At $d\ll w$,  relation (\ref{4}) leads to
\begin{eqnarray} \label{5}
m\approx \frac{2d}{\pi w}.
\end{eqnarray}
The upper parts of the lateral surfaces, ($x=\pm w$, $0\le y\le d/2$), have the following parametric representation ($1\le |u| \le 1/\sqrt{1-m}$):
\begin{eqnarray} \label{6}
\frac{2y}{d}=\frac{f(s(u),m)}{f(1,m)},
\end{eqnarray}
where
\begin{eqnarray} \label{7}
s(u)=\sqrt{\frac{1-(1-m)u^2}{m}}.
\end{eqnarray}
The values $u=\pm 1/\sqrt{1-m}$ correspond to the
equatorial points of the strip, $(\pm w,0)$. For the thin strip (when $d/w\ll 1$ and $m\ll 1$), formula (\ref{6}) can be represented in the explicit form \cite{jetp13}:
\begin{eqnarray} \label{8}
\frac{y}{d}\approx \frac{1}{\pi}[\arcsin(s)+s\sqrt{1-s^2}].
\end{eqnarray}
Interestingly, inaccuracy of this formula does not exceed $8\%$ even at $m=1/2$ (i.e., at $d/2w=1$) and decreases with decreasing  $d/2w$. The above functions $x(u)$ and $y(u)$ are shown in Fig.~\ref{fig3}.

The Meissner currents on the upper and lateral
surfaces of the strip (i.e., in the whole interval $-1/\sqrt{1-m}\le u \le 1/\sqrt{1-m}$) are described by the unified formula:
\begin{eqnarray} \label{9}
J_M(u)=\frac{1}{\sqrt{|1-u^2|}}\left( uH_a+ \frac{I}{2\pi w}\frac{f(1,1-m)}{\sqrt{1-m}}\right).
\end{eqnarray}
Formulas (\ref{2})--(\ref{9}) provide the quantitative description (in the parametric form) of the surface Meissner currents in the strip, including its edge regions. These formulas also enable one to calculate the fractions of the total current $I$ that flow on the upper (lower) and on the two lateral surfaces of the strip,
\begin{eqnarray}  \label{10}
I_{\rm R,L}&=&\frac{I}{\pi}\arcsin(\sqrt{m}) \pm \frac{dH_a\sqrt{m}}{f(1,m)}, \\
I_{\rm upper}&=&I_{\rm lower}=\frac{I}{\pi}\arcsin(\sqrt{1-m}), \nonumber
\end{eqnarray}
where the signs plus and minus refer to the currents on the right ($I_{\rm R}$) and left ($I_{\rm L}$) lateral surfaces, respectively. Of course, $I_{\rm L}+I_{\rm R}+I_{\rm upper}+I_{\rm lower}=I$.

Consider now the above formulas in several limiting cases. In the case of the thin strip ($m\ll 1$), for the points on its upper surface ($u^2< 1$) when these points are not too close to the corners ($1-u^2\gg m$, i.e., at $w-x\gg d$), one finds from Eqs.~(\ref{2}) and (\ref{3}) that $x/w \approx u$. Then, with Eq.~(\ref{9}), we arrive at the well known result obtained in the limit of the infinitely thin strip \cite{ZL,LO,eh93,zeldov94}:
\begin{eqnarray} \label{11}
J_M(x,d/2)\approx \frac{1}{
\sqrt{w^2-x^2}}\left(xH_a+\frac{I}{2\pi}\right)\,,
\end{eqnarray}
where we have taken into account that the factor   $f(1,1-m)/\sqrt{1-m} \approx 1$ at $m\ll 1$ (Fig~\ref{fig2}).
On the other hand, near the corners of the thin strip (for $1-u^2\lesssim m$, or equivalently, at $w-|x|\lesssim d$) formula (\ref{2}) can be  rewritten in the explicit form \cite{jetp13},
\begin{eqnarray} \label{12}
\frac{w-x}{w} \approx \frac{m}{2} s (s^2-1)^{1/2}-\frac{m}{2}
\ln(s+\sqrt{s^2-1}),
\end{eqnarray}
where $s(u)$ is still given by Eq.~(\ref{7}). Now $J_M(x,d/2)$ is not described by simple formula (\ref{11}).
In the limit $|1-u^2|\ll m$, i.e., at $l\equiv w-|x|\ll d$, or at $l\equiv (d/2)-|y|\ll d/2$, the surface current diverges like $l^{-1/3}$ near the corners of the strip \cite{Meissner,jetp13}. In this limiting case formulas (\ref{2})-(\ref{4}) and (\ref{9}) lead to the expression
\begin{eqnarray} \label{13}
J_M\!\approx\!H_a\!\left(\!\pm 1\!+\!\frac{i_H}{2}\right)\! \left(\!\frac{(1-m)d}{6\sqrt{m}f(1,m)\,l}\! \right)^{\!1/3}\!\!\!\!\!\!,~~
\end{eqnarray}
which is valid for the strip of an arbitrary thickness. Here
  \begin{equation} \label{14}
 i_H\equiv \frac{I}{\pi w H_a}\frac{f(1,1-m)}{\sqrt{1-m}},
  \end{equation}
and the signs $\pm$ correspond to the right and left corners, respectively. For the thin strips, expression (\ref{13}) is further simplified since $i_H\approx I/\pi wH_a$ and $f(1,m)\approx \pi m/4$ at $m\ll 1$. The divergence of the current in Eq.~(\ref{13}) should be cut off at $l\lesssim \lambda$, and the current density $j$ throughout the corner region ($w-\lambda \le x \le w$, $(d/2)-\lambda \le |y|\le d/2$) is approximately constant, $j_{\rm crn}(x,y)\sim J_M(x=w-\lambda)/\lambda$. In particular, in the case of the thin strip we obtain
\begin{eqnarray} \label{15}
j_{\rm crn}\!\sim\!\frac{H_a}{\lambda\sqrt{m} }\!\left(\pm 1+\! \frac{i_H}{2} \right)\!\!\left(\frac{2d}{3\pi \lambda} \right)^{\!1/3}\!\!\!\!\!\!\!\!.~~
\end{eqnarray}
Finally, consider the case of a narrow  slab carrying the transport current $I$ in the magnetic field $H_a$ parallel to its surface. This case corresponds to $d \gg 2w$ (i.e, to  $1-m\ll 1$). For this slab, when the coordinate $y$ of a point on a lateral surface is not close to the corners (i.e., at $|u|\gg 1$ and $1-s(u)\gg 1-m$), one finds from Eqs.~(\ref{3}) and Eq.~(\ref{6}) that $2y/d\approx s(u)$, whereas Eqs.~(\ref{9}), (\ref{4}), and (\ref{7}) give
\begin{eqnarray} \label{16}
J_M(y)= \pm H_a+ \frac{I}{2\pi\sqrt{(d/2)^2-y^2}},
\end{eqnarray}
where the estimate $f(1,m)/\sqrt{m} \approx 1$ have been  taken into account again for $m\to 1$, and the signs $\pm$ refer to the right and left lateral surfaces, respectively. Note that the distribution of the current over the surfaces is not uniform. This result is due to the fact that the distributions of the  transport current are identical in the Meissner states of the thin strip and of the narrow slab with the same aspect ratio, and so the distribution in Eq.~(\ref{16}) agrees with that in formula (\ref{11}). On the other hand, the applied magnetic field $H_a$ generates the well-known uniform surface sheet currents $\pm H_a$.

\subsection{Strip with vortex dome}

Using the results of Appendix \ref{A}, one can find the surface sheet currents generated by a vortex dome $B_y(u_0)$ located between points $u_0^{(1)}$ and $u_0^{(2)}$, $u_0^{(1)}\le u_0 \le u_0^{(2)}$, on the upper (and lower) surface of the strip where $-1\le u_0^{(1)}\le u_0^{(2)}\le 1$, and $B_y(u_0)$ is the magnetic induction at the point $u_0$. These currents on the upper surface ($-1\le u \le 1$) have the following form:
 \begin{equation} \label{17}
J_v(u)=\frac{1}{\pi}\int_{u_0^{(1)}}^{u_0^{(2)}} \frac{du_0B_y(u_0) \sqrt{1-u_0^2}}{\mu_0(u_0-u)\sqrt{1-u^2}},
 \end{equation}
whereas on the lateral surfaces ($1\le |u| \le 1/\sqrt{1-m}$) they look like
 \begin{equation}\label{18}
J_v(u)=\frac{1}{\pi}\int_{u_0^{(1)}}^{u_0^{(2)}} \frac{du_0B_y(u_0) \sqrt{1-u_0^2}}{\mu_0(u_0-u)\sqrt{u^2-1}}.
 \end{equation}
Note that similarly to Eq.~(\ref{9}), these currents are, in fact, described by the unified formula in the whole interval $-1/\sqrt{1-m}\le u \le 1/\sqrt{1-m}$.

For the vortex dome to be immobile in the sample, the total sheet current on the upper (lower) surface of the strip, $J_M(u)+J_v(u)$, has to vanish inside the dome, i.e., at $u_0^{(1)} \le u \le u_0^{(2)}$,
 \begin{equation} \label{19}
 J_M(u)+J_v(u)=0.
 \end{equation}
With formulas (\ref{9}) and (\ref{17}), this condition is an integral equation in $B_y(u_0)$ that is solvable analytically \cite{Mus} (see also Appendix~\ref{B} where more general equation (\ref{8b}) is solved with the methods of Ref.~\cite{Mus}). Its solution can be readily written since this equation is formally close to that discussed by Benkraouda and Clem \cite{ben2} if one replaces our variable $u$ by the variable $x$ of Ref.~\cite{ben2}. Eventually we arrive at the following distribution of the magnetic induction  $B_y(u_0)$ describing the static vortex dome on the upper (lower) surface of the sample:
\begin{equation} \label{20}
B_y(u_0^{(1)}\!\!\le u_0\! \le\! u_0^
{(2)}\!)\!=\!\mu_0H_a\frac{\sqrt{\!(u_0^{(2)}-u_0)(u_0- u_0^{(1)})}}{\sqrt{1-u_0^2}},~~~~~~
\end{equation}
where the boundaries $u_0^{(1)}$ and $u_0^{(2)}$ of the dome are not arbitrary. They satisfy the relationship that is the necessary for   this solution to exist \cite{Mus},
 \begin{equation} \label{21}
 u_0^{(1)}+ u_0^{(2)}=-i_H,
 \end{equation}
where the parameter $i_H$ is defined by Eq.~(\ref{14}).
Of course, this static dome exists if $-1\le u_0^{(1)}\le u_0^{(2)}\le 1$. These inequalities impose restrictions on possible values of $I$ and $H_a$. In particular, if the parameter $i_H$
exceeds $2$, the static vortex dome cannot occur in the strip. Note also that there is an arbitrariness in choosing a value of one of  $u_0^{(1)}$ and $u_0^{(2)}$. This fact is a reflection of the dependency of the vortex state on a history of its creation in the sample. For example, if with increasing $H_a$, the vortex dome appears as a result of vortex penetration into the strip, a certain condition on the sheet current  should be fulfilled at the right edge of the sample where the vortices enter the strip  (see Ref.~\cite{jetp13} and also below). This condition leads to an additional equation in $u_0^{(1)}$ and $u_0^{(2)}$, and then these parameters are determined unambiguously. On the other hand, if the dome has already existed in the sample, and the field $H_a$ begins to decrease, the dome expands, but its total magnetic flux has to be constant till the vortex dome reaches one of the edges. The constancy of the flux is another example of the additional condition on $u_0^{(1)}$ and $u_0^{(2)}$.

\begin{figure}[t]
 \centering  \vspace{+9 pt}
\includegraphics[scale=.48]{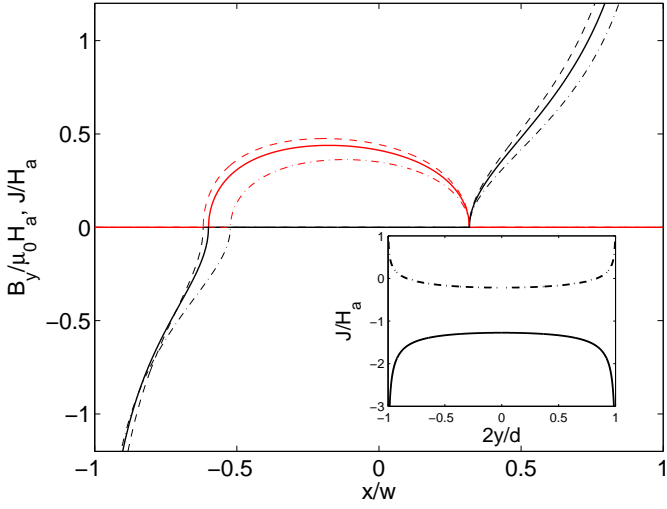}
 \caption{\label{fig4} The $x$-dependences of the magnetic induction  $B_y$ and of the surface sheet current $J$ on the upper surface of the strip, Eqs.~(\ref{2}) and (\ref{20})--(\ref{23}), at $I/(\pi wH_a)=0.3$, and $a_2\equiv x(u_0^{(2)})/w \approx 0.32$. The solid,   dashed, and dash-and-dot lines correspond to $m=0.1$ ($2w/d\approx 10.65$), to the infinitely thin strip ($m$ and $d/w\to 0$), and to $m=1/2$ ($d/2w=1$), respectively. For this $a_2$, formula (\ref{2}) leads to $u_0^{(2)}=0.3$ if $m=0.1$ and to $u_0^{(2)}\approx 0.27$ if $m=1/2$. Inset: The $y$-dependences of the surface sheet current $J$ on the left lateral surface of the strip with $m=0.5$ at the same values of $I/(\pi wH_a)$ and $a_2$ as in the main plot (the solid line), Eqs.~(\ref{6}) and (\ref{24}). For comparison, the dash-and-dot line depicts the Meissner currents on the left lateral surface of this strip without the dome, Eq.~(\ref{9}), at $I/(\pi wH_a)=4$ (i.e., when $2<i_H\approx 2.4<2/\sqrt{1-m}\approx 2.83$).}
 \end{figure}   

With Eqs.~(\ref{9}), (\ref{17}), (\ref{18}), and (\ref{20}), one can calculate the net surface sheet currents outside the vortex dome. Ultimately we find that the sheet currents flowing on the upper surface of the strip in the regions $u_0^{(2)} \le u \le 1$ and $-1\le u \le u_0^{(1)}$ are equal to
\begin{equation} \label{22}
J(u)=H_a\frac{\sqrt{(u-u_0^{(2)})(u- u_0^{(1)})}}{\sqrt{1-u^2}},
\end{equation}
and
\begin{equation}\label{23}
J(u)=-H_a\frac{\sqrt{(u_0^{(2)}-u)( u_0^{(1)}-u)}}{\sqrt{1-u^2}},
\end{equation}
respectively [and $J(u)=0$  at $u_0^{(1)} \le u \le u_0^{(2)}$  according to Eq.~(\ref{19})]. The currents flowing on the lateral surfaces of the strip ($1 \le |u| \le 1/\sqrt{1-m}$) are
\begin{equation} \label{24}
J(u)=\pm H_a\frac{\sqrt{(u-u_0^{(2)})(u-u_0^{(1)})}}{\sqrt{u^2-1}},
\end{equation}
where the signs plus and minus refer to the positive and negative $u$, respectively (i.e., to the right and left lateral surfaces).

\begin{figure}[t]
 \centering  \vspace{+9 pt}
\includegraphics[scale=.48]{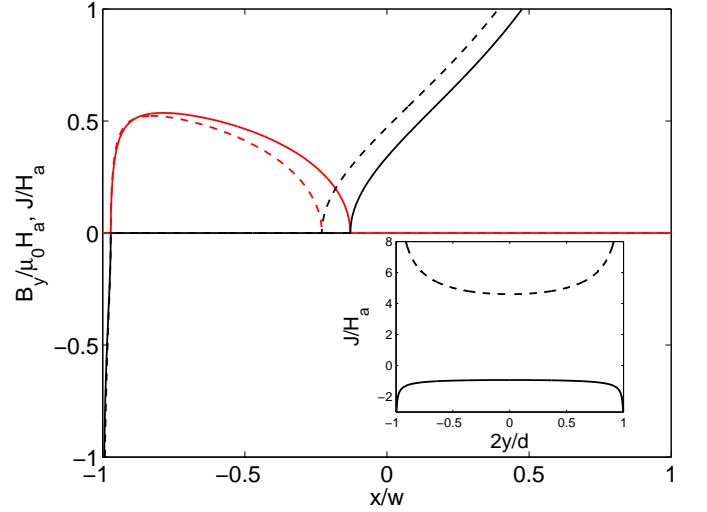}
 \caption{\label{fig5} The $x$-dependences of the magnetic induction  $B_y$ and of the surface sheet current $J$ on the upper surface of the strip, Eqs.~(\ref{2}) and (\ref{20})--(\ref{23}), at $I/(\pi wH_a)=1.2$, and $a_1\equiv x(u_0^{(1)})/w \approx -0.97$ (this  $a_1$ corresponds to $u_0^{(1)}=-0.95$ if $m=0.1$). The solid and dashed lines depict the cases of $m=0.1$ and of the infinitely thin strip ($m\to 0$), respectively. Inset: The $y$-dependences of the surface sheet current $J$ on the left (the solid line) and the right (the dashed line)  lateral surfaces of the strip with $m=0.1$, Eqs.~(\ref{6}) and (\ref{24}); the parameters $I/(\pi wH_a)$ and $a_1$ are the same as in the main plot. }
 \end{figure}   

The $I$-dependence of $J(u)$ in Eqs.~(\ref{22})-(\ref{24}) is implicitly contained in the parameters $u_0^{(1)}$ and $u_0^{(2)}$, and at $u_0^{(2)}-u_0^{(1)}\to 0$, expressions (\ref{22}) and (\ref{24}) transforms into Eq.~(\ref{9}). Indeed, the numerator in these expressions can be rewritten as follows:
\begin{eqnarray*}
(u-u_0^{(2)})(u- u_0^{(1)})=\left(u+\frac{i_H}{2}\right)^2- \frac{(\Delta u)^2}{4},
 \end{eqnarray*}
where $\Delta u\equiv u_0^{(2)}-u_0^{(1)}$ is the width of the dome, and relationship (\ref{21}) has been used to express the position $(u_0^{(2)}+u_0^{(1)})/2$ of its center in terms of $i_H$. If the dome becomes small ($\Delta u \to 0$), one can neglect $(\Delta u)^2/4$, and formulas (\ref{22}) and (\ref{24}) reduce to Eq.~(\ref{9}). If the center of this dome tends to the left edge of the strip ($i_H\to 2$), the vortex state crosses over to the Meissner state.

Although Eq.~(\ref{20}) for the magnetic induction and Eqs.~(\ref{22}) and (\ref{23}) for the currents look like the appropriate expressions for the infinitely thin strip ($d/w \to 0$) \cite{ZL}, formulas (\ref{2}), (\ref{6}), and (\ref{20})--(\ref{24}) describe (in the parametric form) the sheet currents flowing on all the surfaces of the strip with an arbitrary aspect ratio $2w/d$.
These formulas reveal that the difference in the magnetic inductions and the currents for the thin ($d/w\ll  1$) and infinitely thin ($d/w\to 0$) strips can be small everywhere except the  regions near the edges of the sample. Near the edges, these sheet currents are essentially different since they are characterized by the distinct types of their  divergence at $x\to \pm w$; compare Eqs.~(\ref{11}) and (\ref{13}). Figures \ref{fig4} and \ref{fig5} demonstrate the profiles of the magnetic induction and of the sheet current on the upper surfaces of the strips with various aspect ratios. The profiles are plotted at fixed values of $I/(\pi w H_a)$ and one of the vortex-dome boundaries. Interestingly, the differences between the profiles calculated at $m\to 0$ and at finite $m$ are not large even if $m$ approaches $1/2$ (Fig.~\ref{fig4}), and these differences depend on the position of the vortex dome (compare Figs.~\ref{fig4} and \ref{fig5}). The inset in Fig.~\ref{fig4} shows the change of the sheet current on the left lateral surface of the strip when the Meissner state in the sample transforms into the state with the vortex dome on its upper surface. Note also that the sheet-current profile on this lateral surface becomes flat when the left boundary of the dome tends to the left corner of the sample (the inset in Fig.~\ref{fig5}).

It was implied in the above formulas that the vortices filling the dome are straight lines. However, simple considerations (Appendix \ref{B}) show that these vortices should be slightly curved in thin samples. Nevertheless, the analysis presented in Appendices \ref{B} and \ref{C} reveals that the vortex-line curvature has a small effect on the distributions of the currents and the magnetic induction in thin strips at $H_a < H_{c1}$, and we neglect this curvature below.

\begin{figure}[t]
 \centering  \vspace{+9 pt}
\includegraphics[scale=.46]{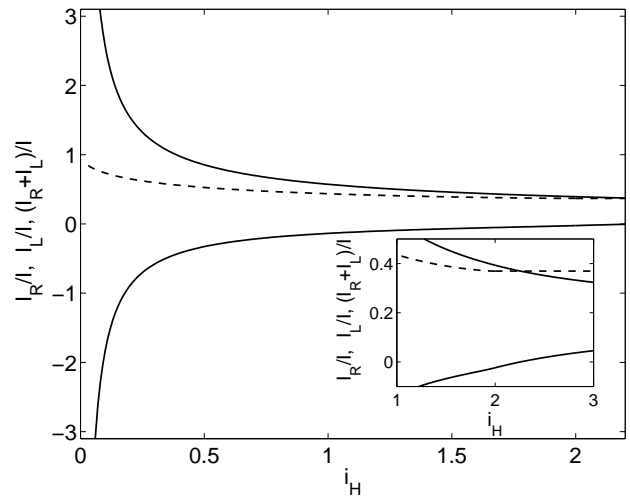}
 \caption{\label{fig6} Dependences of the fractions $I_R/I$, $I_L/I$ (the upper and lower solid lines, respectively) and of  $(I_R+I_L)/I$ (the dashed line) on the parameter $i_H$ given by formula (\ref{14}); $m=0.3$, i.e., $2w/d\approx 2.5$. At $i_H\le 2$ when the vortex dome exists in the sample, these fractions are calculated with formulas (\ref{6}), (\ref{7}), (\ref{21}), and (\ref{24}), assuming  $u_0^{(1)}=-1$. At $i_H\ge 2$ when the vortex dome is absent, these fractions are found with formulas (\ref{10}). The inset: The part of the plot in the enlarged scale.  }
 \end{figure}

Using Eqs.~(\ref{21})--(\ref{24}), one can calculate the total currents flowing on each of the surfaces of the sample. (They are expressible in terms of the elliptic integral.) We present the appropriate formulas only for the thin strips (in the leading order in $m\ll 1$) and when $u_0^{(1)}=-1$,
 \begin{eqnarray}\label{25}
 I_L\!&\approx&\!-H_ad\sqrt{1-\frac{i_H}{2}}, \ \ \ \  \frac{I_L}{I}\approx -\frac{m}{2i_H}\sqrt{1-\frac{i_H}{2}}, \\
 I_R\!&\approx&\! H_ad\, \Phi\!\left(\!\frac{2i_H}{m}\!\right),\ \ \ \ \ \
 \frac{I_R}{I}\approx\frac{m}{2 i_H} \, \Phi\!\left(\!\frac{2i_H}{m}\!\right), \label{26}
 \end{eqnarray}
where $i_H\approx I/\pi wH_a$,
 \[
 \Phi(x)\equiv  \frac{2(1+x)}{\pi}
 \arcsin\!\left(\!\frac{1}{\sqrt{1+x}\!}\right)+ \frac{2\sqrt{x}}{\pi},
 \]
and $I_R$ and $I_L$ are the total currents flowing on the right and left lateral surfaces of the sample, respectively. Interestingly, according to these formulas, at $i_H=m/2$ one has $I_L/I\approx -1$, $I_R/I\approx 1+(2/\pi)$, and $(I_L+I_R)/I\approx 2/\pi$, i.e, more than half of the applied current $I$ flows on the lateral surfaces of the strip at this $i_H$. In the case when $m$ is not small, the fractions $I_L/I$ and $I_R/I$ of the total current $I$ versus the parameter $i_H$ are shown in Fig.~\ref{fig6}. Note that these fractions are continuous functions of $i_H$ at the point  $i_H=2$ below which the vortex dome exists in the sample.

In the next section we shall need the current density in the immediate vicinity of the corners of the thin strips, $j_{\rm crn}$. With Eqs.~(\ref{22})--(\ref{24}), the appropriate expression is obtained much as  Eq.~(\ref{15}),
\begin{eqnarray} \label{27}
j_{\rm crn}\!\sim\!\frac{H_a}{\lambda\sqrt{m} }\! \sqrt{(u_0^{(2)}\pm 1)(u_0^{(1)}\pm 1)}\!\left(\frac{2d}{3\pi \lambda} \right)^{\!1/3}\!\!\!\!\!\!\!\!,
\end{eqnarray}
where the signs $+$ and $-$ correspond to the left and right corners of the strip, respectively. With Eq.~(\ref{21}), it is not difficult to show that the current density $j_{\rm crn}$ at the right corners is always larger than $j_{\rm crn}$ at the left corners if $i_H>0$.

\section{Conditions determining the critical current} \label{III}

When the current $I$ reaches its critical value $I_c$, vortices begin  to cross the sample, entering the strip at its right edge and leaving  it at the left lateral surface. The two types of this process are possible. If the ratio $i_H<2$, the vortex dome exists in the sample. In this case a vortex penetrating into the sample arrives at the right edge of the dome, whereas the left edge of the dome emits another vortex that leaves the strip.  The requirements of the entrance and exit of the vortices impose two additional conditions on the parameters of the dome. One of the conditions together with Eq.~(\ref{21}) unambiguously specify the boundaries $u_0^{(1)}$ and $u_0^{(2)}$ of the vortex dome, and the second one gives the value of $I_c$ at a given $H_a$. If $i_H>2$, the dome is absent, and the vortex crosses the Meissner state of the strip. In this case the distribution of the surface currents, Eq.~(\ref{9}), does not contain undefined parameters, and the critical current $I_c$ is found from the vortex-entry condition at the right edge of the strip. We now discuss the vortex-entry and vortex-exit conditions, and in the next section we find the dependence $I_c(H_a)$.

\subsection{Vortex-entry condition}

For the case of the Meissner state in the strip with $I=0$ and $H_a\neq 0$, the vortex-entry condition was analyzed in Ref.~\cite{jetp13}. Consider now this condition in the general case when both $I$ and $H_a$ are different from zero and when the vortex dome can exist in the sample.

\subsubsection{Bean-Livingston barrier} \label{BL}

Since the currents are maximum at the corners of the sample, it is favorable for a vortex to penetrate into the strip through these points. A small circular vortex arc appearing in one of the corners  overcomes the Bean-Livingston barrier and begins to expand when the current density in the right corners reaches the value $j_0$ \cite{jetp13},
 \begin{equation}\label{28}
j_0\approx \frac{0.92H_{c1}\kappa}{\lambda \ln\kappa},
 \end{equation}
where $H_{c1}=\Phi_0\ln\kappa/(4\pi\mu_0\lambda^2)$ is the lower critical field, $\Phi_0$ is the flux quantum, and $\kappa$ is the Ginzburg-Landau parameter. This $j_0$ is of the order of the depairing current density $j_{dp}$, whereas $j_0\lambda$, the local surface field near the corner, reaches the value of the thermodynamic critical field in the agreement with the results of Refs.~\cite{dezhen,samokh2,galaiko,genen}. Equating this $j_0$ with the current density $j_{\rm crn}$ defined by Eqs.~(\ref{15}) or (\ref{27}), we find the vortex-entry condition  at which the Bean-Livingston barrier disappears for the vortex penetrating through a corner of the sample.

In the case of the Meissner state, Eq.~(\ref{15}) and the requirement $j_{\rm crn}=j_0$ give the following  vortex-entry condition at the right edge of the strip:
 \begin{equation}\label{29}
\frac{H_a}{\sqrt{m}}\left(1+\frac{i_H}{2}\right)
= \frac{0.92\kappa H_{c1}}{\ln\kappa}\!\left(\!\frac{3\pi\! \lambda}{2d}\! \right)^{\!1/3}\!\!\!\!\!\!\!.
 \end{equation}

When the vortex dome exists in the sample, Eqs.~(\ref{27})   and the equality $j_{\rm crn}=j_0$ lead to the condition on $u_0^{(1)}$ and $u_0^{(2)}$, which is additional to Eq.~(\ref{21}),
\begin{eqnarray} \label{30}
\frac{H_a}{\sqrt{m} }\sqrt{\!(1-u_0^{(2)})(1-u_0^{(1)})}= \frac{0.92\kappa H_{c1}}{\ln\kappa}\!\left(\!\frac{3\pi\! \lambda}{2d}\! \right)^{\!1/3}\!\!\!\!\!\!\!.~~~~
\end{eqnarray}

\begin{figure}[t] 
 \centering  \vspace{+9 pt}
\includegraphics[bburx=720,bbury=480,scale=.41]{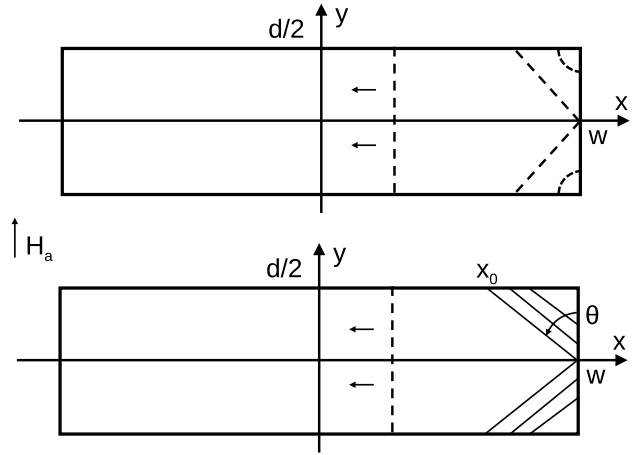}
 \caption{\label{fig7} Two scenarios of the vortex penetration into the strip. Top: $p>p_c$, the Bean-Livingston barrier prevails over the geometrical one. Bottom: $p<p_c$, the penetration of vortices is mainly determined by the geometrical barrier. The parameter $p$ is defined by Eq.~(\ref{44}),  $p_c\approx 0.52$. The dash lines schematically show mobile vortices in the strip, whereas the solid lines inside the strip designate the immobile vortices that are in the equilibrium. These inclined vortices form the flux-line domes on the right lateral surface of the strip.
 } \end{figure}   

\subsubsection{Geometrical barrier}
\label{GB}

Consider now the vortex-entry condition  caused exclusively by the
geometric barrier in the strip, neglecting the attraction of
vortices to the surfaces of the strip. In this case a penetrating
vortex can move towards the left edge of the sample only when its two inclined rectilinear segments meet at the right equatorial point ($x=w$, $y=0$), see Fig.~\ref{fig7}. Consider a vortex which ends at the point $x_0$ of the upper plane of the strip and at the point $y_0=0$ of its lateral surface. The balance between the line tension of the vortex and the forces generated by the surface currents leads to the following equations for $x_0$ and $\theta$ \cite{jetp13}:
\begin{eqnarray} \label{31}
\Phi_0J(x_0,d/2)&=&e_0\sin\theta,  \\
\Phi_0J(w,0)&=&e_0\cos\theta,  \label{32}
\end{eqnarray}
where the sheet currents $J(x,y)$ are determined by the formulas
of the preceding section, $e_0=\Phi_0^2\ln\kappa/(4\pi\mu_0\lambda^2)$ is the line energy of a vortex in an isotropic superconductor, and $\theta <\pi/2$ is the tilt angle of the vortex relative to the lateral surface of the strip. There is also a geometrical relationship between $x_0$ and $\theta$, which is evident from Fig.~\ref{fig7}:
\begin{eqnarray} \label{33}
w-x_0=\frac{d}{2}\tan\theta.
\end{eqnarray}
The three  equations (\ref{31})--(\ref{33}) completely determine the two quantity $\theta$ and $x_0$ and also give the vortex-entry condition in the case of the geometrical barrier.

Let us rewrite these equations using formulas of Sec.~\ref{II}. In the case of the Meissner state ($i_H>2$), Eqs.~(\ref{31}) and (\ref{32}) can be transformed with formulas (\ref{9}) and (\ref{7}) into the form:
\begin{eqnarray}\label{34}
 \cos\theta\!&=&\!\!\frac{H_a}{H_{c1}\sqrt{m}}\left(\!\!1+ \frac{i_H}{2}\sqrt{1\!-\!m}\right), \\
 \sin\theta\!&=&\!\!\frac{H_a}{H_{c1}\sqrt{m}\sqrt{s_0^2-1}}\left( \!\!\sqrt{1\!-\!ms_0^2}+\frac{i_H}{2}\sqrt{1\!-\!m}\right)\!\!,~~ \label{35}
 \end{eqnarray}
where the parameter $s_0$ corresponds to the point
$x_0$ according to Eq.~(\ref{12}). With the use of formulas (\ref{12}), the geometrical relationship  (\ref{33}) looks like
\begin{eqnarray}\label{36}
\tan\theta=\frac{mw}{d}\Big[s_0\sqrt{s_0^2-1}-\ln(s_0+ \sqrt{s_0^2-1})\Big].
 \end{eqnarray}

In the case of the vortex dome in the strip ($i_H<2$), equations Eqs.~(\ref{31}) and (\ref{32}) are rewritten as follows:
\begin{eqnarray}\label{37}
 \cos\theta\!&=&\!\!\frac{H_a}{H_{c1}\sqrt{m}} \sqrt{(1\!-\!\sqrt{1\!-\!m}u_2^{(0)})(1\!-\!\sqrt{1\!-\!m} u_1^{(0)}\!)},~~ \\
 \sin\theta\!&=&\!\!\frac{H_a }{H_{c1}\sqrt{m}\sqrt{s_0^2-1}}\,F,   \label{38}
 \end{eqnarray}
where we have introduced the notation
 \[
F\!\!=\!\sqrt{\!(\sqrt{1\!-\!ms_0^2}\!-\!\sqrt{1\!-\!m}u_2^{(0)}) (\sqrt{1\!-\!ms_0^2}\!-\!\sqrt{1\!-\!m} u_1^{(0)}\!)}.
 \]
As to the geometrical relationship (\ref{36}), it remains unchanged in this case. Note that at $i_H=2$ when only the values $u_1^{(0)}=u_2^{(0)}=-1$ are admissible [see Eq.~(\ref{21})], equations (\ref{37}) and (\ref{38}) cross over to Eqs.~(\ref{34}) and (\ref{35}), respectively.

\begin{figure}[t]
 \centering  \vspace{+9 pt}
\includegraphics[scale=.47]{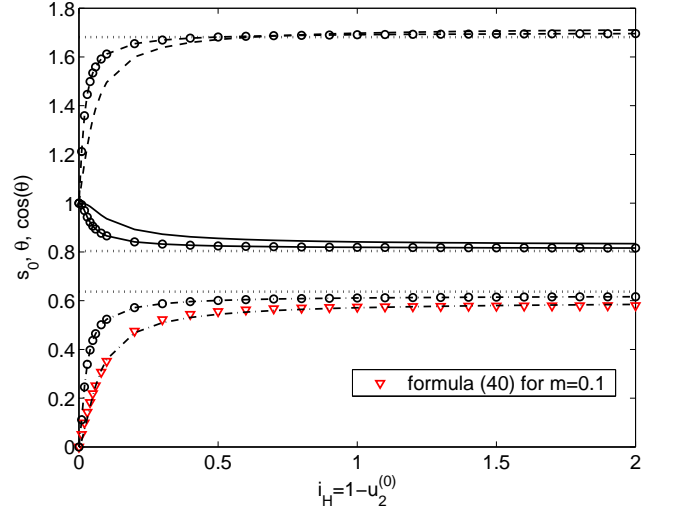}
 \caption{\label{fig8} Dependences of $s_0$ (the dashed lines), $\theta$ (the dot-and-dash lines), and $\cos\theta$ (the solid lines) on the position of the right boundary of the vortex dome $u_2^{(0)}$ at $u_1^{(0)}=-1$ for $m=0.1$ (the lines without circles) and $m=0.04$ (the lines with the circles). The dependences are calculated with Eqs.~(\ref{36})--(\ref{38}) and (\ref{21}). The dotted lines depict the values $s_0\approx 1.68$, $\theta\approx 36.5^{\circ}$, $\cos\theta \approx 0.80$ obtained in the limit $m\to 0$ (see the text). The triangles mark $\theta(i_H,m=0.1)$ calculated with formula (\ref{40}).
 } \end{figure}

Equations (\ref{34})--(\ref{38}) can be solved as follows:
Equating the ratio of  formulas (\ref{35}) and (\ref{34}) to the right hand side of the geometrical condition (\ref{36}), one arrives at the equation determining $s_0$  as a function of $i_H\ge 2$ and of  $m$. Then, formula (\ref{36}) gives $\theta(i_H,m)$, and Eq.~(\ref{34}) becomes the vortex-entry condition which relates $H_a$ to $i_H$ (i.e. to the current) in the case of the Meissner state. Analogously, equating the ratio of  formulas (\ref{38}) and (\ref{37}) to the right hand side of  expression (\ref{36}), one arrives at the equation determining $s_0$  as a functions of $u_1^{(0)}$, $u_2^{(0)}=-i_H-u_1^{(0)}$, and $m$. The angle $\theta$ is obtained from Eq.~(\ref{36}) and $s_0(u_1^{(0)},u_2^{(0)},m)$.
Then, Eq.~(\ref{37}) together with relationship (\ref{21}) leads to the following vortex-entry condition for the strip with the vortex dome:
 \begin{eqnarray}\label{39}
 \frac{H_a}{\sqrt{m}} \sqrt{(1+\frac{i_H}{2}\sqrt{1-m})^2-\!(1-m)\!\left(\!\frac{u_2^{(0)} -u_1^{(0)}}{2}\!\right)^2} \nonumber \\
  = H_{c1}\cos\theta.~~~
 \end{eqnarray}
As it will be clear below, only the case $u_1^{(0)}=-1$ is important for the critical-current calculations. In this case, the dependences of $s_0$ and $\theta$ on $u_2^{(0)}$ (i.e., on $i_H$) are shown in Fig.~\ref{fig8} for two small values of $m$. Interestingly, this figure also demonstrate that at small $m$, the angle $\theta(i_H,m)$ can be well approximated by the function,
 \begin{eqnarray}\label{40}
 \theta\approx (0.637-0.5m)\tanh\left( \frac{0.9i_H}{m+0.3i_H} \right).
 \end{eqnarray}
This approximation of $\theta$ enables one to avoid solving the set of Eqs.~(\ref{36})--(\ref{38}) when the vortex-entry condition (\ref{39}) is used to calculate the critical current.

It should be emphasized that although we deal with the thin strips here, Eqs.~(\ref{34})--(\ref{38}) have been written without recourse to the condition $m\ll 1$. Let us now simplify these equations using the smallness of the parameter $m$. In this case we may put $\sqrt{1-ms_0^2}\approx 1$ since $s_0\sim 1$. Then, the ratio of Eqs.~(\ref{35}) and (\ref{34}) or Eqs.~(\ref{38}) and (\ref{37}) yields
 \[
 \tan\theta\approx \frac{1}{\sqrt{s_0^2-1}}.
 \]
Inserting this formula into Eq.~(\ref{36}), we arrive at the equation in $s_0$ which is independent of $i_H$, $u_2^{(0)}-u_1^{(0)}$, and of the parameter $m$ since $mw/d\approx 2/\pi$ according to Eq.~(\ref{5}). The solution of this equation  gives $s_0\approx 1.68$ and hence $\theta\approx 0.637$ ($36.5^{\circ}$), $\cos\theta \approx 0.80$ \cite{jetp13}. Then, the vortex-entry condition in the case of the Meissner state, Eq.~(\ref{34}), reduces to
 \begin{eqnarray}\label{41}
 \frac{1}{\sqrt{m}}\left(\!H_a+ \frac{I}{2\pi w}\right)\approx 0.80 H_{c1},
 \end{eqnarray}
where we have omitted the factor $f(1,1-m)\approx 1$ in the definition (\ref{14}) of $i_H$. Similarly, the vortex-entry  condition in the case of the vortex dome, Eq.~(\ref{39}), takes the form,
\begin{eqnarray}\label{42}
 \frac{H_a}{\sqrt{m}} \sqrt{(1-u_2^{(0)})(1- u_1^{(0)})} \approx 0.80 H_{c1}.
 \end{eqnarray}
However, it is necessary to keep in mind that condition (\ref{42}) becomes inaccurate if the boundary of the vortex dome $u_2^{(0)}$ is close to the right edge of the strip, $1-u_2^{(0)}\lesssim m$, i.e., if this boundary is at a distance of the order of or less than $d$ from the right lateral surface of the strip. This situation occurs at $H_a\sim H_{c1}$.  In this case one cannot use the approximation $\sqrt{1-m}\approx \sqrt{1-ms_0^2}\approx 1$ everywhere in Eqs.~(\ref{37}) and (\ref{38}). The fail of this approximation is also seen from the dependence of the angle $\theta$ on $u_2^{(0)}$ shown in Fig.~\ref{fig8}. At small $m$, the vortex-entry condition that is valid for all $u_2^{(0)}$ takes the form
 \begin{eqnarray}\label{43}
 \frac{H_a}{\sqrt{m}} \sqrt{\!\left(1+\frac{i_H}{2}\right)^{\!2}\!\!\!-\!(1-m)\! \left(\!\!\frac{u_2^{(0)} -u_1^{(0)}}{2}\!\!\right)^{\!2}}\!\!= H_{c1}\!\cos\theta,~~~
 \end{eqnarray}
where $\theta$ is given by formula (\ref{40}) when $u_1^{(0)}=-1$.

\subsubsection{Two scenarios of the vortex penetration} \label{IIIc}

A comparison of formulas (\ref{29}) and (\ref{41}) or (\ref{30}) and (\ref{42}) shows that for the case of thin strips ($m\ll 1$), the formulas differ in their right hand sides only. The ratio of these right hand sides can be written as $p/p_c$ where the
parameter $p$ is defined as follows:
\begin{equation}\label{44}
p\equiv \frac{\kappa}{\ln\kappa }\left (\frac{\lambda}{d}\right
)^{1/3},
\end{equation}
and $p_c=(0.80/0.92) [2/(3\pi)]^{1/3} \approx 0.52$. Since the
parameter $p$ can be greater or less than its critical value $p_c$, two scenarios of the vortex penetration into the sample are possible. In Ref.~\cite{jetp13} these scenarios were described for the case when the transport current is absent, $I=0$. In this case conditions (\ref{29}) and (\ref{41}) give the vortex penetration fields $H_p^{BL}$ and $H_p^{GB}$ determined by the Bean-Livingston and geometrical barriers, respectively. If $p>p_c$, one has
$H_p^{BL}> H_p^{GB}$, and the true penetration field $H_p$ coincides with $H_p^{BL}$,
\begin{eqnarray}\label{45}
H_p^{BL}\approx 0.80H_{c1}\sqrt{m}\,\frac{p}{p_c}.
\end{eqnarray}
In this case, small vortex segments appearing at the corners of the
strip at $H_a= H_p^{BL}$ immediately expand, merge at the
equatorial point ($x=w$, $y=0$), and the created vortex moves towards the center of the sample, Fig.~\ref{fig7}. This type of the
penetration occurs because at $p>p_c$ and $H_a=H_p^{BL}$, when the
current density in the corner region is close to the depairing
current density, the surface current at the equatorial
point is larger than $0.8H_{c1}$, and the vortex segment
cannot be in the equilibrium at this point.

If the parameter $p$ is less than the critical value $p_c$, one
has $H_p^{BL}< H_p^{GB}$, and the vortex penetration is a
two-stage process. The current density in the vicinity of the
corners reaches the depairing value at $H_a=H_p^{BL}$. At this
field a penetrating vortex enters the sample through the
corner, but it cannot reach the equatorial point since $J(x=w, y=0)$ is less than $0.8H_{c1}$, and so this vortex line will ``hang'' between the corner and the equatorial point $(x=w, y=0)$. With increasing $H_a$, two domes filled by these inclined vortex  lines will expand in the lateral surface of the strip. The penetration field $H_p$ is determined by the condition that the boundaries of these domes meet at the equatorial point, and this field  $H_p$ can be estimated from Eq.~(\ref{41}),
\begin{eqnarray}\label{46}
H_p^{GB}\approx 0.80H_{c1}\sqrt m.
\end{eqnarray}
However, formula (\ref{41}) has been derived, considering a single inclined vortex. Since the vortex domes on the lateral surface of the strip modify the current distribution in the sample, the $H_p$ has to be calculated, taking into account the currents generated by the domes of the inclined vortices.
Nevertheless, as was shown in Ref.~\cite{jetp13}, the maximal decrease of $H_p$ associated with these domes does not exceed $20\%$ as compare to Eq.~(\ref{46}), and we shall neglect this decrease in our subsequent analysis.

Of course, the two described scenarios of the vortex penetration also take place both in the Meissner state with $I\neq 0$ and in the state with  the vortex dome inside the strip if the vortex-entry condition can be  described by Eq.~(\ref{42}). However, as was mentioned above,  condition (\ref{42}) fails if $H_a$ approaches $H_{c1}$. In this situation, the interplay between the Bean-Livingston and geometrical barriers will be discussed in the Sec.~\ref{IV}.

\subsection{Vortex-exit condition}\label{IIIB}

\begin{figure}[t] 
 \centering  \vspace{+9 pt}
\includegraphics[scale=.45]{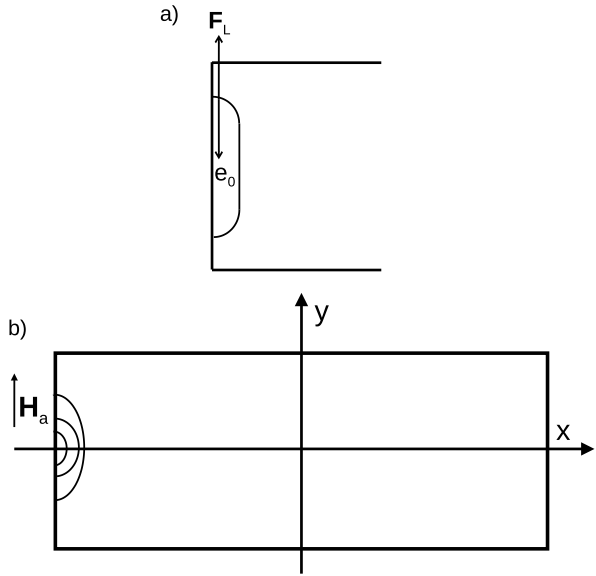}
 \caption{\label{fig9} a) A vortex leaving the left lateral surface of the strip at $2< i_H< 2/\sqrt{1-m}$; $F_L$ is the Lorentz force, and $e_0$ is the force that contracts the vortex. These forces are applied to the vortex segment inside the surface layer of the thickness $\lambda$. b) The unusual vortex state on the left lateral surface of the strip. The state appears if condition (\ref{47}) fails.
   }
 \end{figure}   

Consider now the vortex-exit condition.
If the parameter $i_H$ defined by Eq.~(\ref{14}) exceeds $2/\sqrt{1-m}$, the vortex dome cannot exist in the sample, and the Meissner currents, Eq.~(\ref{9}), are positive on all the surfaces of the strip. Then, if a vortex enters the strip, it  crosses the sample and is expelled from it by the currents on the left lateral surface. If $2< i_H< 2/\sqrt{1-m}$, the vortex dome on the upper surface of the strip is still absent, but the Meissner current, Eq.~(\ref{9}), changes its sign on the left lateral surface (see the inset in Fig.~\ref{fig4}). In this case, if a vortex enters the sample, it is displaced by the positive currents on the upper and lower planes of the strip to the left lateral surface. However only the ends of this vortex leave the sample in the regions near the planes where the currents are positive. A behavior of a vortex segment near the equatorial point $(x=-w,y=0)$ depends on the relation between the Lorentz force $F_L=\Phi_0|J(-w,y)|$ generated by the negative currents on the left  lateral surface of the strip and the force that contracts the vortex and that is equal to its line energy $e_0= \Phi_0^2\ln \kappa/(4\pi \mu_0 \lambda^2)$ (Fig.~\ref{fig9}). The vortex leaves the sample if this $e_0$ exceeds the maximal value $\Phi_0|J(-w,0)|$ of the Lorentz force which stretches the vortex. This condition together with Eq.~(\ref{9}) leads to the restriction on $H_a$,
\begin{equation} \label{47}
\frac{H_a}{\sqrt{m}}\left(1-\frac{i_H\sqrt{1-m}}{2}\right)\le H_{c1},
\end{equation}
where $H_{c1}=\Phi_0\ln\kappa/(4\pi\mu_0\lambda^2)$ is the low critical field. This restriction has a simple physical meaning. The left hand side of Eq.~(\ref{47}) is the magnetic field $H_y$ at the equatorial point $(-w,0)$ outside the strip, and hence this restriction means that a vortex can be  expelled from the superconductor when the local external magnetic field near its surface is less than $H_{c1}$. If the applied magnetic field $H_a$ is less than or of the order of $H_{c1}$, this condition is always  fulfilled. However in samples with perfect surfaces, the Meissner state can occur at $H_a$ exceeding $H_{c1}$ since a vortex cannot enter the strip through its surface because of the Bean-Livingston barrier. In this case an unusual vortex state can, in principle, appear on the left lateral surface of the strip (Fig.~\ref{fig9}).  We return to a discussion of this issue in Sec.~\ref{V}.

If $i_H< 2$ (and vortices penetrate into the sample), the vortex dome described in the preceding section  appears on the upper surface of the strip. When $u_0^{(1)}>-1$, the negative currents flow on the upper (lower) surface in the interval $-1<u< u_0^{(1)}$ [see Eq.~(\ref{23}) and Figs.~\ref{fig4}, \ref{fig5}]. These currents prevent vortices of the dome from leaving the sample. Thus, the necessary condition of the vortex exit is
\begin{equation} \label{48}
u_0^{(1)}=-1.
\end{equation}
Under this condition, the left boundary of the dome touches the left lateral surface of the strip. In this case, according to Eq.~(\ref{24}), the sheet currents on the left lateral surface $J(-w,y)$ are approximately equal to $-H_a\sqrt{1-0.5i_H}$, and one may expect that the vortices will easily leave the sample at least for $H_a<H_{c1}$; see also Appendix \ref{C}.

\section{Critical current of thin strips} \label{IV}

\subsection{Zero magnetic field}\label{IVa}

Using the results of the preceding section, we now calculate the critical current $I_c(H_a)$ of the thin strips. Let us begin with the case of zero applied magnetic field $H_a$. In this case one has $i_H\approx I/\pi w H_a\to \infty$, and the vortex dome is absent on the upper surface of the sample. Then, the vortex-entry conditions (\ref{29}) and (\ref{41})  give the following expressions for $I_c(0)$ in the cases of the Bean-Livingston and geometrical barriers, respectively:
 \begin{eqnarray}\label{49}
I_c^{\rm BL}(0)\!&\approx& 1.60\pi w \sqrt{m}\frac{p}{p_c}H_{c1}, \\
I_c^{\rm GB}(0)\!&\approx &\!1.60\pi w \sqrt{m}H_{c1},
\label{50}
 \end{eqnarray}
where  we have used formula (\ref{44}). Inserting formula (\ref{5}) for $m$ into Eq.~(\ref{50}), we arrive at the expression
 \[
 I_c^{\rm GB}(0)\approx 1.60\sqrt{2\pi dw}H_{c1}.
 \]
Note that the appropriate result of Ref.~\cite{ben2} (see the Introduction) differs from this expression only by the factor which is close to unity. However, the true critical current $I_c(0)$ coincides with the largest value of $I_c^{\rm GB}(0)$ and $I_c^{\rm BL}(0)$.

If $I_c^{\rm GB}(0)> I_c^{\rm BL}(0)$ (i.e. $p<p_c\approx 0.52$), and if the current $I$ in an experiment increases from zero value, the segments of vortices begin to enter the strip through its
corners at $I=I_c^{\rm BL}(0)$, but they cannot reach the equatorial points. Hence, the domes of the inclined vortices appear on the
lateral surfaces of the strip. The situation is similar to that described in Sec.~\ref{IIIc}, but the inclined vortices on the left and right lateral surfaces now have opposite vorticities.
The critical value of $I$ is reached only  when the domes touch each other at the equatorial points, Fig.~\ref{fig7}. These domes, as was mentioned in Sec.~\ref{IIIc}, generate additional surface currents and reduce the critical current $I_c(0)$ as compared to $I_c^{\rm GB}(0)$ given above. Although the reduction increases with decreasing $p/p_c$, it  does not exceed $20\%$ according to the results of Ref.~\cite{jetp13}, and hence the above $I_c^{\rm GB}(0)$ can be a reasonable estimate of the critical current. In the opposite case, when $I_c^{\rm GB}(0)< I_c^{\rm BL}(0)$ (i.e. $p>p_c$), the domes of the inclined vortices do not appear in the strip, and $I_c(0)= I_c^{\rm BL}(0)$, Eq.~(\ref{49}). Note that this $I_c^{\rm BL}(0)$ for the thin samples ($m\ll 1$) can be also rewritten as follows:
 \begin{eqnarray}\label{51}
I_c^{\rm BL}(0)\!\approx 2.18\pi w \sqrt{m}H_c\!\left(\frac{\lambda}{d}\right)^{\!1/3}\!\!\!\!\!\!\!\!\!
\approx 2.18\sqrt{2\pi d w}H_c\!\left(
\frac{\lambda}{d}\right)^{\!1/3}\!\!\!\!\!\!\!\!,~~~~
 \end{eqnarray}
where $H_c=\sqrt{2}\kappa H_{c1}/\ln\kappa$ is thermodynamic critical field, and we have used formulas (\ref{5}), (\ref{44}), and the value $p_c\approx 0.52$ in obtaining Eq.~(\ref{51}).

We are now in position to trace the increasing role of the Bean-Livingston barrier with decreasing $d$ (see the Introduction). If a sample is so thick (as compared to $\lambda$) that $p<p_c$, the geometrical barrier determines $I_c(0)$. In this case  $I_c(0)=I_c^{\rm GB}(0)\propto \sqrt d$ according to Eq.~(\ref{50}). With decreasing $d$, at $d_{\rm cr}= \lambda [\kappa/(0.52\ln\kappa)]^3> \lambda$, the crossover in the $d$-dependence of $I_c(0)$ occurs since at $d\le d_{\rm cr}$, one has $p>p_c$, the Bean-Livingston barrier begins to govern the critical current, and $I_c(0)$ is determined by expression (\ref{51}). When $d\to \lambda$, this expression agrees with formulas of Refs.~\cite{likh,mak-c,plourde,LO,aslam,bezug} for $I_c(0)$ within a factor of the order of unity.

Interestingly, the simple relation between the critical current $I_c(0)$ and the vortex-penetration field $H_p$ measured at zero $I$,
 \begin{eqnarray}\label{52}
I_c(0)=2\pi w H_p,
 \end{eqnarray}
follows both from Eqs.~(\ref{45}), (\ref{49}) in the case of the Bean-Livingston barrier and from formulas (\ref{46}), (\ref{50}) for the case of the geometrical barrier. However, for the samples with $p<p_c$, the domes of the inclined vortices on the lateral surfaces of the strip, in principle, can modify this relation. When the critical current  is measured in such samples at $H_a=0$, the vortex domes on the right and left lateral surfaces have opposite vorticities,  whereas at the measurement of the vortex-penetration field at zero current, the domes are of the same vorticity on both these surfaces. Nevertheless, since the vortex domes located on one of the lateral surfaces of a thin strip generate relatively weak currents on its other surface, relation (\ref{52}) is likely to remain true at least approximately.

\subsection{Critical current and the Bean-Livingston barrier}

\begin{figure}[tbh] 
 \centering  \vspace{+9 pt}
\includegraphics[scale=.48]{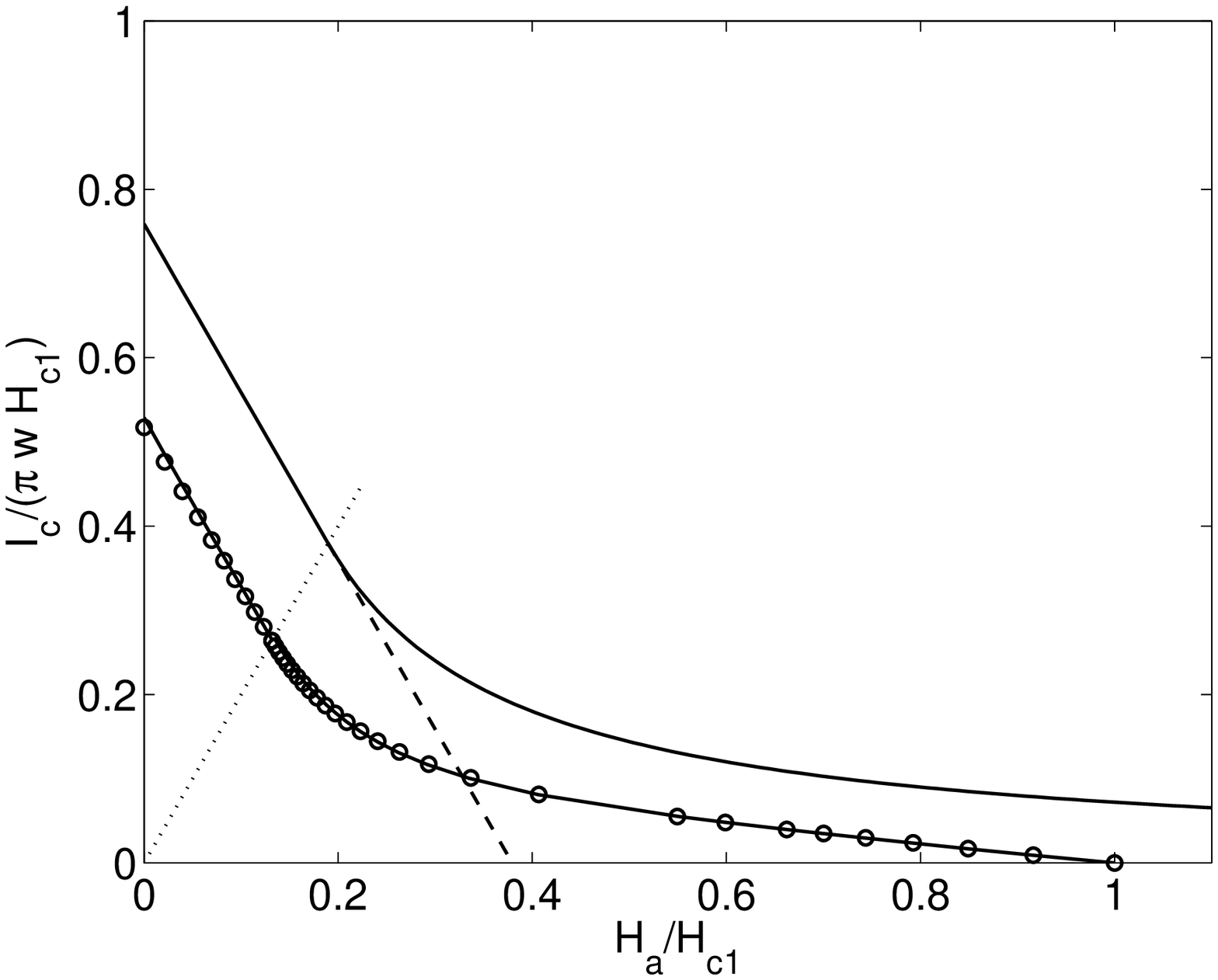}
 \caption{\label{fig10} Dependences $I_c^{\rm BL}(H_a)$ [the upper solid line, Eqs.~(\ref{49}), (\ref{52})--(\ref{54})] and $I_c^{\rm GB}(H_a)$ [the lower line, Eqs.~(\ref{55})-(\ref{57}), and (\ref{40})] plotted for $m=0.1$ and $p/p_c=1.5$. The dotted line corresponds to the condition $i_H=2$; the circles show $I_c^{\rm GB}(H_a)$ calculated with Eqs.~(\ref{55})-(\ref{57}) and with the solution of Eqs.~(\ref{34})-(\ref{38}) for $\theta$ instead of Eq.~(\ref{40}). The true critical current $I_c$ coincides with $I_c^{\rm BL}$. The straight dashed line is the continuation of $I_c(H_a)$ shown in the region $i_H\ge 2$; the intercept produced by this line in the $H_a$ axis gives the vortex-penetration field $H_p=H_p^{\rm BL}$.
  } \end{figure}

If $p>p_c$ and hence if $I_c(0)$ is determined by the Bean-Livingston barrier, the vortex-entry condition (\ref{29}) enables one to find the $H_a$-dependence of the critical current $I_c$ in the Meissner state (i.e, at $i_H\ge 2$) of the thin strips,
 \begin{eqnarray}\label{53}
I_c^{\rm BL}(H_a)=I_c^{\rm BL}(0)\left(1-\frac{H_a}{2H_*^{\rm BL}}\right),\ \ \ H_a\le H_*^{\rm BL},
 \end{eqnarray}
where $H_*^{\rm BL}=H_p^{\rm BL}/2$, and the penetration field $H_p^{\rm BL}$ is determined by formula (\ref{45}). At $H_a=H_*^{\rm BL}$ the ratio $i_H$ reaches the value $2$, and the vortex dome appears in the sample. Taking into account the vortex-exit condition (\ref{48}) and inserting $u_2^{(2)}=1-i_H$ into the vortex-entry relationship (\ref{30}), we arrive at the equations in $i_H$ for the case of the strip with the vortex dome,
 \begin{eqnarray*}
H_a\sqrt{2i_H} = 0.8\sqrt{m}H_{c1}\frac{p}{p_c},\ \ \ H_a \ge H_*^{\rm BL}.
 \end{eqnarray*}
This equation with the definition $i_H\approx I/\pi w H_a$ and formulas (\ref{45}), (\ref{49}) gives
 \begin{eqnarray}\label{54}
I_c^{\rm BL}(H_a)=I_c^{\rm BL}(0)\frac{H_*^{\rm BL}}{2H_a},\ \ \ H_a \ge H_*^{\rm BL}.
 \end{eqnarray}
Thus, at $p>p_c$ when the Bean-Livingston barrier governs the critical current, the $H_a$-dependence of this current  is determined by formulas (\ref{53}) and (\ref{54}), Fig.~\ref{fig10}. This dependence is similar to that obtained in Ref.~\cite{ben2}, but  $I_c(0)$ and $H_p$ are now defined by the other formulas which are characteristic of the Bean-Livingston barrier.

Since the critical current $I_c^{\rm BL}(H_a)$ for thin strips is described by simple explicit formulas (\ref{53}), (\ref{54}), we can easily trace its dependence on the aspect ratio $d/2w \ll 1$ at unchangeable other parameters. In particular, we may assume that the width $2w$ increases at a fixed thickness $d$, which defines the parameter $p$ in Eq.~(\ref{44}). According to formulas (\ref{51}) and (\ref{52}), we have  $I_c^{\rm BL}(0)\propto w^{1/2}$, $H_*^{\rm BL}\propto w^{-1/2}$, and the coefficient before $H_a$ in formula (\ref{53}) is equal to $2\pi w$. On the other hand, the critical current $I_c^{\rm BL}(H_a)$ described by Eq.~(\ref{54}) is independent of $w$.

We now describe the evolution of the vortex state in the sample in the process of the critical-current measurements when at given $H_a$  the current $I$ gradually increases in the strip with $p>p_c$. If $H_a<H_*^{\rm BL}=H_p^{\rm BL}/2$, vortices begin to penetrate into the sample through its right lateral surface only at the critical current $I_c(H_a)=I_c^{\rm BL}(H_a)$. They cross the strip, and the stationary vortex dome does not appear on its upper (lower) surface. If $H_*^{\rm BL}< H_a< H_p^{\rm BL}$, the sample is still in the Meissner state at $I=0$. When $I$ increases, the vortex penetration into the sample starts when $I$ reaches the dashed line in Fig.~\ref{fig10}, and a vortex dome like that in Fig.~\ref{fig4}  appears inside the sample. This dome does not touch the left lateral surface, and the penetrating vortices do not cross the sample. With increasing $I$, they accumulate in the dome, which gradually expands and shifts to the left until its  boundary touches the left lateral surface of the strip at $I=I_c^{\rm BL}(H_a)$. The situation just before the touching is illustrated by Fig.~\ref{fig5}. If $H_a> H_p^{\rm BL}$, the vortex dome exists in the sample at $I=0$, and it look like the dome in Fig.~\ref{fig4} (with its center being at $x=0$).  With increasing $I$, the dome shifts to the left and expands, much as in the case $H_*^{\rm BL}< H_a< H_p^{\rm BL}$.
Note that for $p>p_c$, the line $I=I_c^{\rm GB}(H_a)$ does not manifest itself experimentally.

\subsection{Critical current and the geometrical barrier}\label{IVc}

\begin{figure}[tbh] 
 \centering  \vspace{+9 pt}
\includegraphics[scale=.48]{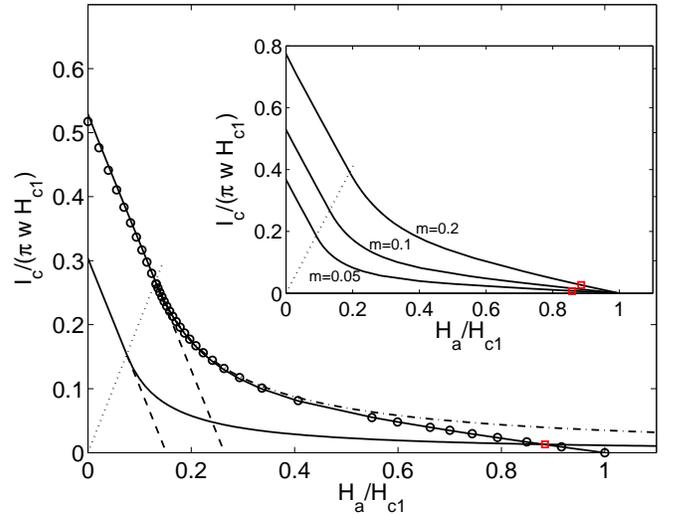}
 \caption{\label{fig11} Dependences $I_c^{\rm BL}(H_a)$ [the lower solid line] and $I_c^{\rm GB}(H_a)$ [the upper solid line] for $m=0.1$ and $p/p_c=0.6$. The figure is similar to Fig.~\ref{fig10}  [in particular, the circles show $I_c^{\rm GB}(H_a)$ calculated with Eqs.~(\ref{55})-(\ref{57}) and with the solution of Eqs.~(\ref{34})-(\ref{38}) for $\theta$]. However, there is a crossing point (marked by the square) of  $I_c^{\rm BL}(H_a)$ and $I_c^{\rm GB}(H_a)$ at which the crossover in $I_c(H_a)$ occurs. At $H_a$ below this point, $I_c(H_a)=I_c^{\rm GB}(H_a)$, whereas $I_c(H_a)=I_c^{\rm BL}(H_a)$ above this point. The dot-and-dash line depicts the function described by formula (\ref{59}) \cite{ben2}. The intercepts produced by straight dashed lines in the $H_a$ axis give $H_p^{\rm BL}$ and $H_p=H_p^{\rm GB}$.
 Inset: Dependences $I_c^{\rm GB}(H_a)$ for different aspect ratios of the strip ($m=0.05, 0.1, 0.2$) at the fixed value of $p/p_c=0.6$. The squares mark the crossing points of $I_c^{\rm GB}(H_a)$ and $I_c^{\rm BL}(H_a)$ for $m=0.05$ and $0.2$.
 } \end{figure}

For the thin strips, the $H_a$-dependence of $I_c^{\rm GB}$ can be readily obtained, using vortex-exit condition (\ref{48}) and  representing Eqs.~(\ref{34}), (\ref{43}), and definition (\ref{14}) in the parametric form,
 \begin{eqnarray}\label{55}
 H_a\!&=&\frac{\sqrt{m} H_{c1}\cos\theta}{\sqrt{(1+\frac{i_H}{2})^2-\!(1-m)(1 -\frac{i_H}{2})^2}
  },  \\
   H_a\!&=&\frac{\sqrt{m} H_{c1}\cos\theta}{(1+\frac{i_H}{2})}, \label{56} \\
  I_c^{\rm GB}\!&\approx& \pi w H_a i_H, \label{57}
 \end{eqnarray}
where the angle $\theta=\theta(i_H)$ can be well approximated by the explicit  formula (\ref{40}), and $i_H$ plays the role of the parameter which  runs the intervals $0 \le i_H\le 2$ and $2\le i_H<\infty$  for Eqs.~(\ref{55}) and (\ref{56}), respectively. [In other words, Eq.~(\ref{56}) gives $H_a(i_H)$ when $H_a$ is below the field $0.5\sqrt{m} H_{c1}\cos[\theta(i_H=2)]$, whereas for $H_a$ lying above this field, the function $H_a(i_H)$ is specified by Eq.~(\ref{55}).]
The dependence $I_c^{\rm GB}(H_a)$ thus obtained is shown in Figs.~\ref{fig10} and \ref{fig11}. For comparison, in these figures,  we also depict the dependence $I_c^{\rm GB}(H_a)$ calculated with angle $\theta$ found from the set of Eqs.~(\ref{34})--(\ref{38}).

At the limiting value $i_H=0$, expressions (\ref{55}) and (\ref{40}) give  $H_a= H_{c1}$. If $H_a$ is noticeably less than $H_{c1}$ and hence if $i_H\gg m$, the formulas for $I_c^{\rm GB}(H_a)$ can be further simplified. In this case, one can set $1-m\approx 1$ and $\cos\theta\approx 0.8$ in  Eq.~(\ref{55}). Then, we arrive at the simple  dependence $I_c^{\rm GB}(H_a)$ that is similar to that given by Eqs.~(\ref{53}) and (\ref{54}),
 \begin{eqnarray}\label{58}
I_c^{\rm GB}(H_a)&\approx& I_c^{\rm GB}(0)\left(1- \frac{H_a}{2H_*^{\rm GB}}\right),\ \ \ H_a\le H_*^{\rm GB}\!, \\
I_c^{\rm GB}(H_a) &\approx& I_c^{\rm GB}(0)\frac{H_*^{\rm GB}}{2H_a},\ \ \ \ \ \ \ \ \ \
 \ \ \ \ \ H_a\ge H_*^{\rm GB}\!, \label{59}
 \end{eqnarray}
where $H_*^{\rm GB}=H_p^{\rm GB}/2$, see Eq.~(\ref{46}).
In fact, this dependence coincides with that obtained in Ref.~\cite{ben2}, and we show it in Fig.~\ref{fig11}. It is seen that when $H_a$ approaches $H_{c1}$, the dependence $I_c^{\rm GB}(H_a)$ obtained with Eqs.~(\ref{55})--(\ref{57}) essentially deviates from that derived by Benkraouda and Clem \cite{ben2}.

The true critical current $I_c(H_a)$ is determined by the largest function  of $I_c^{\rm GB}(H_a)$ and $I_c^{\rm BL}(H_a)$.
If $p<p_c$, $I_c(0)$ is determined by the geometrical barrier.
The dependence $I_c^{\rm GB}(H_a)$ presented in Fig.~\ref{fig11} shows that although at $p<p_c$ the geometrical barrier prevails over the Bean-Livingston one in the region of the weak magnetic fields, the crossover  magnetic field $H_{cr}$ necessarily exists above which the Bean-Livingston barrier begins to govern the critical current. This crossover  field corresponds to the crossing point of the functions $I_c^{\rm GB}(H_a)$ and $I_c^{\rm BL}(H_a)$ that is visible in Fig.~\ref{fig11}. Note that the crossing point would be absent if Eq.~(\ref{59}) were used for the description of $I_c^{\rm GB}(H_a)$. It is also worth noting that at the crossing point each dome of the inclined vortices on the right lateral surface of strip reduces to a single vortex line, and so the effect of these domes on the $I_c^{\rm GB}$ that was mentioned in Sec.~\ref{IVa} is negligible near this point.

According to formulas (\ref{50}), (\ref{52}) and (\ref{58}), (\ref{59}), the critical current $I_c^{\rm GB}(H_a)$ depends on $w$ for a fixed value of $d$ in the same way as $I_c^{\rm BL}(H_a)$ if $H_a$ is not close to $H_{c1}$. The inset in Fig.~\ref{fig11} shows $I_c^{\rm GB}(H_a)$ for different aspect ratios $d/2w$ of the strip in the whole field range $0\le H_a\le H_{c1}$.

\begin{figure}[t]
 \centering  \vspace{+9 pt}
\includegraphics[scale=.48]{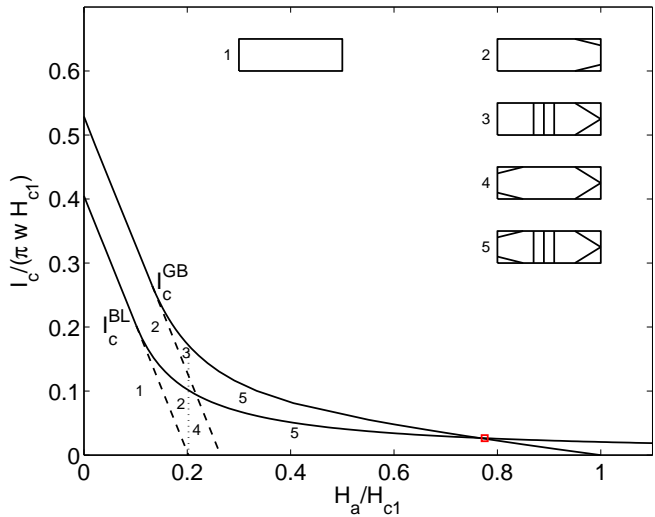}
 \caption{\label{fig11a} Dependences $I_c^{\rm BL}(H_a)$ [the lower solid line] and $I_c^{\rm GB}(H_a)$ [the upper solid line] for $m=0.1$ and $p/p_c=0.8$. The figure is similar to Fig.~\ref{fig11}. The intercepts produced by straight dashed lines in the $H_a$ axis give $H_p^{\rm BL}$ and $H_p=H_p^{\rm GB}$.
 The numbers from $1$ to $5$ mark the regions in the $H_a$-$I$ plane (at $p/p_c<0.5$, the region $3$ is absent). In these regions, different vortex states occur in the strip when the current increases from zero to $I_c$ at a fixed $H_a$. These states are schematically shown in the figure (in particular, $1$ denotes the  Meissner state). Note that each dome of the inclined vortices on the lateral surfaces of the strip is depicted as a single inclined line.
 } \end{figure}   

In the case $p<p_c$, an evolution of the vortex states in the strip in the process of measuring the critical current can be considered similarly to the case $p>p_c$. Let the current gradually increase at given $H_a$. In the initial state (at $I=0$), the strip is either in the Meissner state if $0<H_a<H_p^{\rm BL}$, or the domes of the inclined vortices exist on both lateral surfaces of the strip when  $H_p^{\rm BL} <H_a <H_p^{\rm GB}$, or both the inclined vortices and the dome on the upper surface are present in the sample if $H_p^{\rm GB}<H_a$. As an example, consider the situation when in the initial state  the field $H_a$ is less than  $H_p^{\rm BL}$ (state $1$ in Fig.~\ref{fig11a}). When $I$ crosses the line $I_c^{\rm BL}(H_a)$ for $H_a<H_*^{\rm BL}$ or its continuation (the lower dashed line) for $H_*^{\rm BL} <H_a <H_p^{\rm BL}$, the inclined vortices appear on the right lateral surface of the strip (state $2$), and the domes of these vortices expand with increasing $I$. When $I$ reaches the critical current $I_c^{\rm GB}(H_a)$ in the region $H_a<H_*^{\rm GB}$ or the upper dashed line in the interval $H_*^{\rm GB} <H_a <H_p^{\rm BL}$, the vortices located on the lateral surface become able to penetrate into the bulk of the strip. They either cross the sample without creation of a dome on the upper plane of the strip in the first case or accumulate in such a dome (state $3$) if $H_*^{\rm GB} <H_a <H_p^{\rm BL}$. In the latter case the dome does not touch the left lateral surface of the strip, and this touch occurs at the critical current $I_c(H_a)=I_c^{\rm GB}(H_a)$. Note that at $p<p_c$ and $I<I_c(H_a)$, the vortex state in the strip  depends not only on the geometrical barrier but also on the Bean-Livingston one even if  $H_a <H_{cr}$.

\section{Discussion}\label{V}

Let us first discuss the constraints that have been assumed in the preceding sections. In this paper we have completely disregarded the bulk vortex pinning. A combined effect of the edge barrier and of the bulk pinning on the critical current of the thin strips with $d< \lambda$ was analyzed by Elistratov et al.~\cite{mak-c}, considering these strips as infinitely thin. Under the same approximation $d/w\to 0$, the results obtained in Ref.~\cite{mak-c} can be readily extended to the case $w\gg d> \lambda$. Then one arrives at the following qualitative conclusions: At $H_a=0$ the bulk pinning is inessential if $j_{cp}d<2H_p\sim \sqrt{m}H_{c1}$ where $j_{cp}$ is the critical current density characterizing the pinning forces in the bulk of the strip, and $H_p$ is the penetration field, Eq.~(\ref{52}). With increasing $H_a$, the effect of the bulk pinning on $I_c$ generally enhances, but in the interval of the interest, $0<H_a\lesssim H_{c1}$, this pinning remains negligible if $j_{cp}d\ll  m H_{c1}$. At $j_{cp}d\sim  m H_{c1}$, the vortex state in the strip is still qualitatively similar to that without the bulk pinning, i.e., there are vortex-free regions and a vortex dome in the sample. In other words, in this case a nonzero $j_{cp}$ leads solely to a  quantitative change in the $H_a$-dependence of the critical current. Only at $j_{cp}d\gtrsim H_{c1}$ may one expect that the distributions of the current and of the magnetic induction in the strip take on the usual Bean critical-state forms. It is clear that if with increasing temperature $T$, the critical current density $j_{cp}(T)$ decreases steeper than $H_{c1}(T)$ (see, e.g.,  Appendix in Ref.~\cite{peak}), there is a region near the temperature $T_c$ of the superconducting transition where the bulk pinning can be neglected. This region can be a sufficiently wide in the high-$T_c$ superconductors in which the thermal fluctuations lead to the thermal depinning \cite{bl}.

So far we have considered the case of the isotropic superconducting strips. An anisotropy of the superconducting material is inessential for the states with straight vertical vortices. For the inclined vortices, which are important for  the analysis of the geometrical barrier, the anisotropy can be taken into account in the same way as in Ref.~\cite{jetp13}, and figure 7 of that paper demonstrates that the anisotropy does not noticeably change the results obtained for the isotropic case. However, the anisotropy should be generally taken into account for the curved vortices that are analyzed in Appendices \ref{B} and \ref{C}.

In Sec.~\ref{III}, the vortex-entry conditions (\ref{29}), (\ref{30}), (\ref{41})--(\ref{43}) were derived under the assumption that a vortex can penetrate into the strip only if the Bean-Livingston and geometrical barriers vanishes. Strictly speaking, the conditions thus obtained are valid at sufficiently low temperatures since at $T>0$, a thermally activated vortex can overcome these barriers even if they are finite. It is clear that this effect is most pronounced for high-$T_c$ superconductors; see, e.g., Ref.~\cite{Kop90}. The  thermal activation decreases the critical current $I_c$ specified by the Bean-Livingston or the geometrical barrier, and this decrease of $I_c$ can be estimated using ideas of Ref.~\cite{bur2} (see also Ref.~\cite{bl}). With the activation, the additional factor $[1+(T/U_{BL})\ln(t/t_0)]^{-1}$ appears in the right hand sides of Eqs.~(\ref{29}), (\ref{30}) where $t$ is the characteristic time of the appropriate experiment, $t_0$ is a certain microscopic time \cite{bur2,bl}, and the characteristic value $U_{BL}$ of the Bean-Livingston barrier is of the order of $e_0\xi$. The similar factor $[1+(T/U_{GB})\ln(t/t_0)]^{-1}$ appears in the right hand sides of Eqs.~(\ref{41})--(\ref{43}), but the characteristic magnitude of the geometrical barrier, $U_{BG}\sim e_0d$, essentially exceeds $U_{BL}$. Therefore, the temperature primarily suppresses the Bean-Livingston barrier at the corners of the strip.  Note that the modification of the right hand sides of Eqs.~(\ref{29}), (\ref{30}) and of Eqs.~(\ref{41}), (\ref{42}) leads to the appearance of the ratio of the above-mentioned two factors in the definition of the  parameter $p$. Hence, if at sufficiently low temperatures when both the factors are close to unity, one has $p>p_c$,  an increase in $T$ will decrease the modified parameter $p(T)$, and a crossover from $I_c=I_c^{\rm BL}$ to $I_c=I_c^{\rm GB}$ will take place at a certain temperature. In other words, one may expect a relative enhancement of  the role of the geometrical barrier in the origin of $I_c$ with increasing temperature.

Let us now discuss the possibility of appearing the unusual vortex state on the left lateral surface of the strip, see Sec.~\ref{IIIB}.  This state can occur when $2/\sqrt{1-m}>i_H>2$, i.e., when there is no vortex dome on the upper surface of the strip, and if inequality (\ref{47}) fails. A failure of Eq.~(\ref{47}) can take place only when the Bean-Livingston barrier prevails over the geometrical one ($p>p_c$).   Assuming the following values of the applied magnetic field $H_a\approx H_p^{\rm BL}/2=\sqrt{m}\cos\theta H_{c1}p/2p_c$ and of the applied current $I\approx I_c(0)/2$ (i.e., $i_H\approx 2$), one finds that the unusual state can appear if
 \begin{eqnarray}\label{60}
  p\equiv \frac{\kappa}{\ln\kappa }\left (\frac{\lambda}{d}\right
)^{1/3}>\frac{2p_c}{(1-\sqrt{1-m})\cos\theta}\,,
 \end{eqnarray}
where $\cos\theta\approx 0.8$ and $p_c\approx 0.52$.
This condition leads  to a restriction on the parameters of the supeconducting material of the strip. For example, for  $\lambda=0.2\,\mu$m,  $d=10\,\mu$m, $2w=110\,\mu$m ($m\approx 0.1$), one obtains the severe restriction on $\kappa$: $\kappa >600$, which mainly results from the small value of $m$. However, using formulas (\ref{4}), (\ref{9}), (\ref{13}), and (\ref{28}), condition (\ref{60}) can be written not only for the thin strips with small values of $m$ but also for a slab with an arbitrary value of this parameter. This enables one to search for the unusual vortex state on the upper plane of the thin strips when the applied field $H_a$ is parallel to the $x$-axis in Fig.~\ref{fig1}. Then, $2w$ plays the role of the thickness of the slab, whereas $d$ is its width, and the new parameter $m$ coinciding with former $1-m$ is close to unity. Eventually, we obtain that at $H_a$ near the field $H_p^{\rm BL}/2\approx \cos\theta H_{c1} p/2p_c$ and at $I\approx I_c(0)/2$ (i.e., at $i_H\approx 2$), the condition for the existence of the  unusual vortex state looks like
 \begin{eqnarray}\label{61}
  p\equiv \frac{\kappa}{\ln\kappa }\left (\frac{\lambda}{d}\right
)^{1/3}&>&\frac{2p_c}{(1-\sqrt{m})\cos\theta} \left(\frac{m(1-m)\pi}{4f(1,m)}\right)^{1/3} \nonumber \\
&\approx& \frac{2p_c}{(1-\sqrt{m})\cos\theta},
 \end{eqnarray}
where $m$, $\cos\theta$ and $p_c$ are the same as in Eq.~(\ref{60}), and we have used the formula $f(1,m\ll 1)\approx \pi m/4$.
With the same values of $\lambda$, $d$, and $2w$ as in Eq.~(\ref{60}), inequality (\ref{61}) leads to $\kappa> 22$. Moreover, if one takes into account that due to the anisotropy of superconducting material, the lower critical field in the plane of the strip, $\varepsilon H_{c1}$, can be less than the field $H_{c1}$ for the direction perpendicular to this plane, the right hand side of  formula (\ref{61}) has to be multiplied by the  anisotropy factor $\varepsilon<1$. In this case the requirement $p>p_c$ may be sufficient for the fulfilment of condition (\ref{61}). It is also worth noting that according to formula (\ref{61}), the applied field $H_a \approx \cos\theta H_{c1} p/2p_c$ exceeds $H_{c1}/(1-\sqrt{m}) \gtrsim H_{c1}$. Therefore, in order to observe the unusual vortex state, it is necessary to exclude the thermally activated penetration of vortices into the sample through its surfaces; the penetration should occur only through its corners. This imply the high quality of the surfaces of the strip and fairly low temperatures of the measurements, i.e., low-$T_c$ superconductors seems to be most suitable for such experiments.

Consider now results of the experimental investigations of the surface barrier in thin NbSe$_2$ \cite{Pal98} and Bi$_2$Sr$_2$CaCu$_2$O$_8$ \cite{Fuchs98,Fuchs98b,Haim09} superconducting strips. In these investigations, distributions of the magnetic field $H_y^{\rm ac}(x)$ induced by the alternating transport current of the magnitude $I_{\rm ac}$ were measured at different temperatures $T$ on the  upper surfaces of superconducting strips placed in a relatively large dc magnetic fields $H^{\rm dc}=H_y^{\rm dc}\gg |H_y^{\rm ac}(x)|$. The obtained profiles $H_y^{\rm ac}(x)$ permitted the authors of the above-mentioned papers to analyze the distributions of the current  $I_{\rm ac}$ across the width of the strips and hence to obtain information on the role of the surface barrier in the vortex dynamics for different temperatures. It was found that in the close vicinity of $T_c$ the current flows uniformly across the sample, and the barrier is negligible in this situation. However, with cooling, a noticeable  fraction of the current concentrates near the edges of the strip, which is the characteristic feature of the surface barrier \cite{Fuchs98}, and this fraction increases with decreasing $T$. When the decreasing $T$ reaches a  temperature $T_x(H^{\rm dc})$, the currents flow only near the edges, and the vortices shift from one side of the sample to the other and back during the ac cycle, but they do not enter and leave the sample, i.e., they become trapped in the strip \cite{Haim09}. It is essential that the bulk vortex pinning begins to play an important role only at a lower temperature $T_d(H^{\rm dc})$ than $T_x(H^{\rm dc})$.

Taking into account the results of the previous section, the above-mention experimental findings can be interpreted as follows: With decreasing temperature, the lower critical field $H_{c1}$ increases, and the point corresponding to ($H^{\rm dc}$, $I_{\rm ac}$) on the $H$--$I$ diagram like Fig.~\ref{fig10} ($p>p_c$ for samples in Refs.~\cite{Pal98,Fuchs98,Fuchs98b,Haim09}) shifts along the straight line, $I/H_a=I_{\rm ac}/ H^{\rm dc}=\,$const., towards the origin. On the other hand, the line $I_c(H_a)$ in the diagram  gradually shifts  due to the $T$-dependence of the parameter $p$, see Eqs.~(\ref{44}), (\ref{45}), (\ref{49}), (\ref{53}), and (\ref{54}).
When the point in the diagram is above the  line $I_c(H_a)$, the flux flow occurs in the sample during a part of the ac cycle when the current exceeds the critical one. In the other part  of the cycle, when the current is less than its critical value, it flows near the edges (outside the vortex dome). The closer the point to the line $I_c(H_a)$, the larger is the duration of the subcritical vortex states for which the current concentrates near the edges. At $T\le T_x(H^{\rm dc})$ one has $I_{\rm ac}\le I_c(H^{\rm dc},T)$, and so the vortices do not cross the sample, i.e., they are trapped in the strip. If the critical current is described by Eq.~(\ref{54}), we obtain the relationship between the $T_x$ and $H^{\rm dc}$, which follows from Eqs.~(\ref{44}), (\ref{45}), (\ref{49}), (\ref{54}),
 \begin{eqnarray}\label{62}
H^{\rm dc}&=&\frac{I_c^{\rm BL}(0)H_*^{\rm BL}}{2I_{ac}}\propto (H_{c1}p)^2\propto [\lambda(T_x)]^{-n} \nonumber \\
&\propto& \left(1-\frac{T_x^2}{T_c^2}\right)^{n/2},
 \end{eqnarray}
where $p\propto \lambda^{1/3}$ and $n=10/3$ if the flux creep is neglected, and for definiteness, we have assumed here that $[\lambda(T)]^{-2}\propto 1-(T/T_c)^2$. With the thermal activation of vortices, which is essential for Bi$_2$Sr$_2$CaCu$_2$O$_8$, one should also take into account the above-mentioned dependence of the parameter $p$ on $T$ associated with the creep. Moreover,  in the experiments, the field $H^{\rm dc}$ noticeably exceeds $H_{c1}(T_x)$.
As explained in Appendix \ref{C}, for a magnetic field $H_a$ of the order of $H_{c1}$, the curved vortex at the right boundary of the dome touches the right lateral surface of the strip. Then, at larger values of $H_a$ the inclined vortices will exist on this surface even though the Bean-Livingston barrier prevails over the geometrical one. These inclined vortices may modify the dependence $I_c(H_a)$ given by Eq.~(\ref{54}) since they, in general, change the vortex-entry condition (\ref{30}). Thus, the experimental results of Refs.~\cite{Pal98,Fuchs98,Fuchs98b,Haim09} actually probe the $H_a$ dependence of the critical current in the range $H_a>H_{c1}$ which has not been considered here.

\section{Conclusions}\label{VI}

The critical current $I_c$ of a thin superconducting strip of the rectangular cross section is calculated in absence of flux-line pinning. The width $2w$ and the thickness $d$ of the strip are assumed to be much larger than the London penetration depth. It is shown that the critical current depends on the parameter $p$, Eq.~(\ref{44}), characterizing the interplay between the Bean-Livingston and geometrical barriers. If this parameter exceeds the critical value $p_c$, the Bean-Livingston barrier in the corners of the strip determines the critical current $I_c(0)$ at zero applied magnetic field $H_a$. Otherwise, at $p<p_c$, the critical current $I_c(0)$ is determined by the geometrical barrier. Since the parameter $p$ increases with decreasing $d$, the crossover in the $d$-dependence of $I_c(0)$ occurs. This crossover explains the increasing role of the Bean-Livingston barrier in $I_c(0)$ with decreasing the thickness $d$ of the sample and answers the question raised in Introduction.

With increasing magnetic field $H_a$, the critical current $I_c(H_a)$ decreases, and the strip with this current is either in the Meissner state or in the state with a stationary vortex dome on its upper (lower) surface. In the latter case, and if $p<p_c$, i.e., if the geometrical barrier prevails over the Bean-Livingston one at zero $H_a$, a crossover in the dependence $I_c(H_a)$ occurs at a magnetic field that is less than $H_{c1}$, Fig.~\ref{fig11}. This crossover is due to fact that the increase in the applied magnetic field enhances the role of Bean-Livingston barrier, and ultimately this barrier prevails over the geometrical one. At $p>p_c$ the Bean-Livingston barrier dominates for all the magnetic fields $H_a<H_{c1}$, Fig.~\ref{fig10}. For the magnetic fields $H_a$ exceeding $H_{c1}$, the vortices fill the whole bulk of the strip, and a dome of the inclined vortices is expected to appear on the lateral surface of the sample where the vortex penetration occurs, Appendix \ref{C}. This dome may change the $H_a$-dependence of the critical current, and this dependence remains to be studied theoretically. Interestingly, experiments like in Refs.~\cite{Pal98,Fuchs98,Fuchs98b,Haim09} can shed light on the critical current in this range of the magnetic fields.

The analysis of the vortex exit from the strip shows that the unusual vortex state can appear on the surface of the sample where vortices leave the strip, see Fig.~\ref{fig9}. This state may occur if  the applied magnetic field $H_a$ is approximately half the vortex-penetration field $H_p$ measured at $I=0$ and if the applied current $I$ is equal to the appropriate critical current $I_c(H_a)$. It is most favorable to observe this state if the magnetic field $H_a$ lies in the plane of the strip (across its width) since in this case its upper (or lower) plane plays the role of the vortex-exit surface.

 \begin{figure}[t] 
 \centering  \vspace{+9 pt}
\includegraphics[bburx=720,bbury=500,scale=0.41]{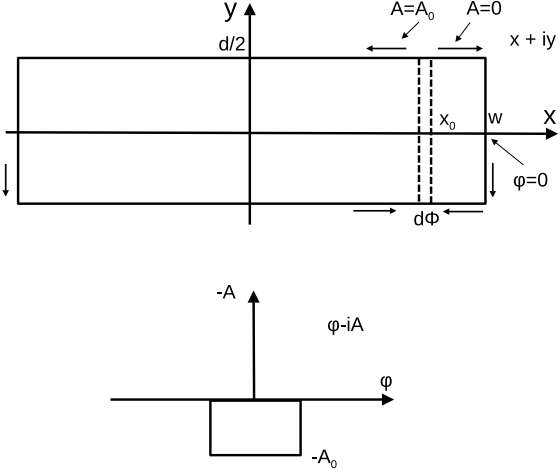}
\caption{\label{fig12} Top: The strip with the thin layer of
vertical vortices piercing its upper and lower planes at the point with the coordinate $x_0$. The boundaries of the layer carrying the flux $d\Phi$ are shown by the dash lines. Arrows indicate directions of the magnetic-field lines $A=0$ and $A=d\Phi/\mu_0\equiv A_0$ that leave and enter the layer and that adjoin the surface of the strip. Bottom: the rectangle in the complex plane $\varphi-iA$. The conformal mapping mentioned in the text  transforms the exterior of the strip in the complex plane $x+iy$ onto the interior of this rectangle.
 } \end{figure}   

\appendix

\section{Currents generated by a ``layer'' of vertical vortices
in the strip.}\label{A}

Consider a thin ``layer'' of the vertical vortices in the strip, Fig.~\ref{fig12} (top). In the $z$ direction, this layer extends to infinity. Let the position of the layer in the $x$ axis be $x_0$, and let this $x_0$ correspond to the parameter $u_0$ ($-1 \le u_0 \le 1$), Eq.~(\ref{2}). The width of the layer is determined by the small interval $du_0$  or by the appropriate  $dx_0$, and the layer carries the magnetic flux $d\Phi=B_y(x_0)dx_0$ where $B_y(x_0)=B_y(u_0)$ is the magnetic induction at the upper surface of the strip at the point $x_0(u_0)$.

The surface currents generated by the layer can be found using results of Sec.~II and a conformal mapping \cite{tfkp} which transforms the exterior of the strip in the complex plane $x+iy$
onto the interior of the rectangle shown in Fig.~\ref{fig12}
(bottom). The lower and upper sides of this rectangle correspond to  the magnetic-field lines $A=0$ and $A=d\Phi/\mu_0$ that adjoin the surface of the strip and that are shown in Fig.~\ref{fig12} (top), whereas the lateral sides of the rectangle correspond to the infinitesimal intervals $dx_0$ carrying the flux $d\Phi$ on the upper and lower planes of the strip. Ultimately we find that at a point $u$, the surface current generated by the layer is given by
\begin{equation}\label{1a}
\!\!\!\!J(u)\!=\!
\frac{B_y(u_0) du_0}{\mu_0\pi(u_0-u)}\!\left (\frac{1-u_0^2}{1-u^2} \right)^{1/2},
 \end{equation}
if $-1\le u\le 1$ (the point $u$ lies on the upper surface of the strip), and by
\begin{equation}\label{2a}
\!\!\!\!J(u)\!=
\frac{B_y(u_0) du_0}{\mu_0\pi(u_0-u)}\!\left (\frac{1-u_0^2}{u^2-
1}\right)^{1/2},
 \end{equation}
if $1\le |u|\le 1/\sqrt{1-m}$ (the point is on the lateral surfaces).

The currents generated by {\it two} vertical vortex layers located symmetrically relative to the $y$ axis were found previously  \cite{jetp13}. The expression that follows from Eqs.~(\ref{1a}) and (\ref{2a}) in the case of the two layers reproduces formula (34) in  Ref.~\cite{jetp13}.

\section{Curvature of flux lines in the strips}\label{B}

It follows from Eq.~(\ref{20}) that the derivative of the magnetic induction, $\partial B_y/\partial x$, differs from zero inside the strip in the region of the vortex dome. However, for this dome to be immobile, the Lorentz force and hence the current density $j_z=(\partial B_y/\partial x -\partial B_x/\partial y)/\mu_0$  have to be equal to zero in this region. This means that $\partial B_x/\partial y$ also differs from zero inside the dome, i.e., the vortices are curved. The curvature of the vortices generates elastic forces applied to the vortex lattice in addition to the Lorentz forces generated by a current. It is known \cite{camp} that the effect of the elasticity of the curved lattice on the balance of the forces can be taken into account if the current density in the condition ${\bf j}=0$ is considered as ${\bf j}={\rm rot}{\bf H}$ rather than ${\bf j}=(1/\mu_0)\,{\rm rot}{\bf B}$ where ${\bf H}$ is the thermodynamic magnetic field, ${\bf H}=\partial F/\partial {\bf B}$, and $F$ is the free-energy density of the superconductor. In  isotropic superconductors, the field ${\bf H}$ is parallel to ${\bf B}$, and for definiteness, we take the function $H(B)$ in the form:
\begin{equation}\label{1b}
H=\sqrt{H_{c1}^2 +(B/\mu_0)^2},
 \end{equation}
which models well the real dependence of $H$ on $B$ at sufficiently low $B<\mu_0H_{c1}$ \cite{clem,eh03,nurit}.
Note that at $B\ll \mu_0H_{c1}$, one has $H\approx H_{c1}$. In this case the elasticity of the vortices plays an important role, their small curvature is able to ``compensates'' the nonzero  $\partial B_y/\partial x$ in the strip, and the vortices in the dome are practically straight lines in the approximation  $H\approx H_{c1}$ \cite{jetp13}.
Below we consider the situation when the curvature of the flux lines  in a strip is small but is not negligible. As it will be clear from subsequent formulas, this situation really occurs in the thin strips, $d\ll w$. Under the assumption of the small curvature, we obtain the following expression describing the shape of the vortex line, $x(y)$, in the thin strips,
 \begin{eqnarray}\label{2b}
 x(y)\approx x(0)+ \frac{y^2}{2H(B)}\frac{d H}{d(B^2)} \frac{\partial B_y^2}{\partial x},
 \end{eqnarray}
where $x(0)$ is a position of this line in the $x$ axis, i.e., at $y=0$, and $\partial B_y^2/\partial x$ is determined by Eqs.~(\ref{2}) and (\ref{20}). Thus, the deflection of the line, $\Delta x\equiv x(d/2)-x(0)$, is equal to
 \begin{eqnarray}\label{3b}
 \frac{2\Delta x}{d}\approx \frac{d}{4H(B)}\frac{d H}{d(B^2)} \frac{\partial B_y^2}{\partial x},
 \end{eqnarray}
and it is generally small for the thin strips, $\Delta x/d \sim (d/w)H_a^2/H^2$. In this estimate of $\Delta x/d$, we have used Eq.~(\ref{1b}) and the expression for $\partial B_y^2/\partial x$ that results from  formulas (\ref{2}) and (\ref{20}):
 \begin{eqnarray} \label{4b}
   \frac{1}{\mu_0^2} \frac{\partial B_y^2}{\partial x}\!&=&\! \frac{H_a^2}{w}\sqrt{1+\frac{mu^2}{1-u^2}} \Big[\frac{u_0^{(1)}+ u_0^{(2)}}{(1-u)^2} \nonumber \\
   &-& 2u\frac{(1+u_0^{(1)})(1+u_0^{(2)})}{(1-u^2)^2}\Big],
 \end{eqnarray}
where we have neglected the factor $[f(1,1-m)/(1-m)]\approx 1$.
It also follows from Eq.~(\ref{2b}) that  the angle $\theta\approx dx(y)/dy$ of the deviation of vortex line from the $y$ axis is small in agreement with above assumption. This angle can become of the order of unity only for the vortices near the boundaries of the dome and only if $H_a\sim H_{c1}$.

The curvature of the vortex lines produces the components $H_x$ and $B_x$ near the upper surface of the thin strip in the region of the vortex dome,
 \begin{eqnarray}\label{5b}
  H_x&\approx& H\frac{dx(y)}{dy}|_{y=d/2}\approx  \frac{d}{2}\frac{d H}{d(B^2)} \frac{\partial B_y^2}{\partial x}, \\
  B_x&\approx& B_y\frac{dx(y)}{dy}|_{y=d/2}\approx  \frac{d}{2H(B)}\frac{d H}{d(B^2)} \frac{\partial B_y^2}{\partial x}B_y. \label{6b}
   \end{eqnarray}
Since $H_x(y)$ is continuous at $y=d/2$ \cite{LL}, formulas (\ref{5b}), (\ref{6b}) reveal the jump in the tangential component of the magnetic induction, $\Delta B_x\equiv \mu_0 H_x- B_x$, at the upper surface of the sample. This jump generates \cite{LL} the surface sheet current $-J_s$ where
\begin{eqnarray}\label{7b}
   J_s=\frac{\Delta B_x}{\mu_0}\approx   \frac{d}{2}\frac{d H}{d(B^2)} \frac{\partial B_y^2}{\partial x}\left(1- \frac{B_y}{\mu_0 H(B)}  \right).
      \end{eqnarray}

\begin{figure}[t]
 \centering  \vspace{+9 pt}
\includegraphics[scale=.47]{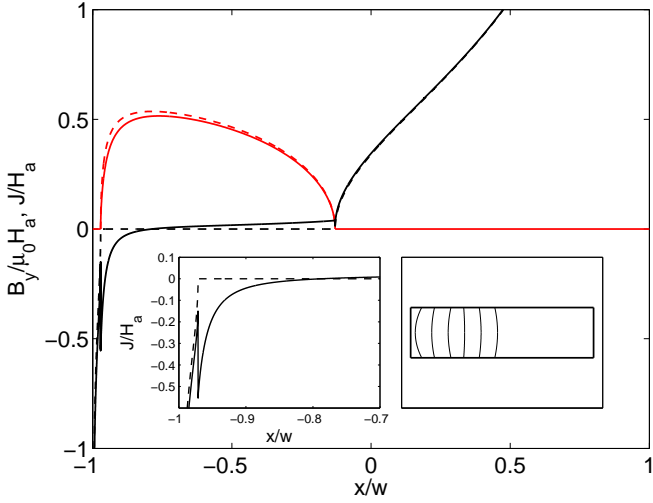}
\caption{\label{fig13} The $x$-dependences of the magnetic induction $B_y$ and of the surface sheet current $J$ on the upper surface of the strip with $m=0.1$. The dashed lines reproduce the vortex dome and the sheet current from Fig.~\ref{fig5} ($u_0^{(1)}=-0.95$, $I/\pi wH_a=1.2$; according to Eq.~(\ref{30}), this vortex state is realized at $H_a/H_{c1}\approx 0.2566$ if $p/p_c=1.5$).  The solid lines show the dome and the sheet current calculated with Eqs.~(\ref{1b}), (\ref{4b}) and (\ref{7b})--(\ref{12b}) for the same values of $u_0^{(1)}$ and $u_0^{(2)}$ as in Fig.~\ref{fig5}. According to Eq.~(\ref{11b}), these values of $u_0^{(i)}$ are realized at $H_a/H_{c1}\approx 0.2566$ and  $I/(\pi wH_a)\approx 1.1975$ if $p/p_c=1.5$. The left inset: The surface sheet current near the left edge of the strip in the enlarged scale. The right inset: Outline of the curved vortices in the strip.
 } \end{figure}

Strictly speaking, the vortex dome has to be found self-consistently from the equation
 \begin{eqnarray}\label{8b}
  J_M(u) + J_v(u)=- J_s(u),
   \end{eqnarray}
which generalizes Eq.~(\ref{19}).
However, for the thin strips considered here, it is sufficient to insert formulas (\ref{20}) and (\ref{4b}) into the right hand side of expression (\ref{7b}) and to solve Eq.~(\ref{8b}) with the $J_s$ thus obtained. In this approximation, the solution for the vortex dome on the upper surface of the strip can be obtained with the methods of Ref.~\cite{Mus}. Ultimately we arrive at
\begin{eqnarray} \label{9b}
\frac{1}{\mu_0}B_y(u_0^{(1)}\!\!\le\!u_0\!\le u_0^
{(2)})\!=\!H_a\frac{R(u_0)}{\sqrt{1-u_0^2}}\left[1+f(u_0) \right],~~~
\end{eqnarray}
where $R(u)\equiv \sqrt{\!(u_0^{(2)}-u)(u- u_0^{(1)})}$, and
 \begin{eqnarray}\label{10b}
f(u_0)=\frac{1}{\pi H_a}
\int_{u_0^{(1)}}^{u_0^{(2)}}\!\!\! \frac{J_s(u)\sqrt{1-u^2}du}{(u-u_0)R(u)}.
\end{eqnarray}
It is necessary to emphasize that $R(u_0)f(u_0)\to 0$  when $u_0$, lying inside the interval $(u_0^{(1)},u_0^{(2)})$, tends to one of its boundaries $u_0^{(1)}$ or $u_0^{(2)}$. The boundaries of the vortex dome, $u_0^{(1)}$ and $u_0^{(2)}$, are related to each other with the equation generalizing formula (\ref{21}),
\begin{eqnarray} \label{11b}
u_0^{(1)}+ u_0^{(2)}+i_H+ \frac{2}{\pi}\int_{u_0^{(1)}}^{u_0^{(2)}}\!\!\! \frac{J_s(u)\sqrt{1-u^2}du}{H_aR(u)}=0.~~~~~
 \end{eqnarray}
The sheet currents outside the vortex dome, $-1/\sqrt{1-m}\le u \le u_0^{(1)}$ and $u_0^{(2)}\le u \le -1/\sqrt{1-m}$, are given by the expression,
\begin{eqnarray} \label{12b}
J(u)\!&=&\!\frac{H_a}{\sqrt{|1-u^2|}}(u+\frac{i_H}{2}) \nonumber \\
&+&\!\frac{1}{\pi\sqrt{|1-u^2|}} \int_{u_0^{(1)}}^{u_0^{(2)}}\!\!\!\frac{B_y(u_0)\sqrt{1- u_0^2}du_0}{\mu_0(u_0-u)} \nonumber \\
&=&\pm H_a\frac{\sqrt{(u-u_0^{(2)})(u- u_0^{(1)})}}{\sqrt{|1-u^2|}}\Big(1+ f(u)\Big),~~~
\end{eqnarray}
where relationship (\ref{11b}) has been used in obtaining the last line; $B_y(u_0)$ is described by Eq.~(\ref{9b}); the signs $\pm$ correspond to $u>u_0^{(2)}$ and $u<u_0^{(1)}$, respectively, and the function $f(u)$ with $u$ lying outside the interval $(u_0^{(1)},u_0^{(2)})$ is still defined by Eq.~(\ref{10b}). However at $u$ approaching $u_0^{(1)}$ from below or at $u$ approaching $u_0^{(2)}$ from above, the function $f(u)$ diverges so that $J(u)\to -J_s(u_0^{(1)})$ or $J(u)\to -J_s(u_0^{(2)})$, respectively. In other words, the sheet current $J(u)$ is continues function of $u$, but in the vicinities of the points $u_0^{(1)}$ and  $u_0^{(2)}$, it changes sharply, see Fig.~\ref{fig13}.
At $J_s=0$, formulas (\ref{9b})--(\ref{12b}) reproduce Eqs.~(\ref{20})--(\ref{24}).

It is clear that the anisotropy of the superconducting material is important for the curved vortices. Let us briefly outline how the anisotropy can be taken into account in the formulas of this Appendix, assuming that a relation between ${\bf H}$ and ${\bf B}$ is known. The anisotropy leads to that the direction of the local magnetic induction ${\bf B}(x,y)$ (i.e., of the vortices inside the strip) defined by the angle $\theta(x,y)$ does not generally coincide with the local magnetic-field direction which we shall describe by the angle $\theta_H(x,y)$. Under the assumption of the small angles $\theta_H$ in the sample, one obtains that the $dy/dx$ calculated with Eq.~(\ref{2b}) still gives the angle $\theta_H$, $dy/dx \approx \theta_H(x,y)$. The angle $\theta(x,y)$ of the magnetic induction  can be found from the relation between ${\bf H}$ and ${\bf B}$, using
 \[
 \sin(\theta- \theta_H) =\frac{[{\bf B}\times {\bf H}]}{BH},
 \]
and this angle is not necessarily small if the anisotropy is strong.
Knowing $\theta_H$ and $\theta$ near the upper surface of the strip ($y=d/2$), one can find the current $J_s$,
 \begin{eqnarray} \label{13b}
J_s\approx (H\theta_H- \frac{B_y}{\mu_0}\tan\theta)|_{y=d/2}.
 \end{eqnarray}
As to formulas (\ref{8b})--(\ref{12b}), they remain unchanged.

\section{Curvature of vortex lines and the vortex-exit and vortex-entry conditions}\label{C}

In the case of the thin strips ($m\ll 1$), if $u$ is  near  the left boundary of the dome, and if this boundary is close to the left lateral surface of the strip ($u\to u_0^{(1)}\approx -1$), expression (\ref{4b}) for $\partial B_y^2/\partial x$ takes the form,
 \begin{eqnarray}\label{1c}
   \frac{1}{\mu_0^2}\frac{\partial B_y^2}{\partial x}&\approx& \!-\frac{H_a^2}{w} \sqrt{1+\frac{m(u_0^{(1)})^2}{1-(u_0^{(1)})^2}}\frac{(2u_0^{(1)} +i_H)}{1-(u_0^{(1)})^2} \nonumber \\
    &\approx& \frac{H_a^2}{2w}\sqrt{1+\frac{m}{2(1+u_0^{(1)})}}\frac{(2 -i_H)}{(1+u_0^{(1)})}.
 \end{eqnarray}
On the other hand, when $1-(u_0^{(1)})^2\ll m \ll 1$, i.e., when $w+a_1\ll d$ where the $x$-coordinate $a_1$ corresponds to $u_0^{(1)}$,  formulas (\ref{5}), (\ref{7}), (\ref{12}) yield,
 \begin{eqnarray}\label{2c}
[1-(u_0^{(1)})^2]^{3/2}\!\approx \frac{3m^{1/2}(w+a_1)}{w}.~~~
 \end{eqnarray}
With formulas (\ref{3b}), (\ref{1c}), and (\ref{2c}), we arrive at the following deflection of the vortex lying at left boundary of the dome:
 \begin{eqnarray}\label{3c}
 \frac{\Delta x}{d}\approx \frac{(2-i_H)\mu_0^2H_a^2}{24H_{c1}}\frac{d H}{d(B^2)} \frac{d}{w+a_1}.
 \end{eqnarray}
When $\Delta x=w+a_1$, the vortex at the left boundary of the dome touches the left equatorial point of the strip. This touch occurs at $a_1$ determined by the expression,
\begin{eqnarray}\label{4c}
 \frac{w+a_1}{d}&\approx& \frac{H_a}{H_{c1}} \sqrt{\frac{(2-i_H)\mu_0^2H_{c1}}{24}\frac{d H}{d(B^2)}} \nonumber \\
 &\approx& \frac{H_a}{H_{c1}} \sqrt{\frac{(2-i_H)}{48}},
 \end{eqnarray}
where, in the last equality, we have used $d H/d(B^2)=1/(2\mu_0^2H_{c1})$ that follows from (\ref{1b}) at $u=u_0^{(1)}$ (i.e., at $B_y\to 0)$.
At this $a_1$, the parameter $u_0^{(1)}$ is equal to
 \begin{eqnarray}\label{5c}
u_0^{(1)}\approx -\sqrt{1-m\left(\frac{9\pi^2 (2-i_H)H_a^2}{192H_{c1}^2}\right)^{1/3}}.
 \end{eqnarray}
It is clear that for thin strips ($m\ll 1$) and at $H_a \le H_{c1}$, formula (\ref{5c}) practically coincides with condition (\ref{48}).

When the curved vortex touches the left equatorial point, a further infinitesimal increase in the applied current seems to trigger the following process: The curved vortex breaks into two segments (one end of each segment is on the left lateral surface of the strip, and the other end is on the upper or lower plane). The line tension contracts these segments since the sheet currents on the lateral surface are  approximately equal to $-H_a\sqrt{1-0.5i_H}$ and are less than $H_{c1}$ at $H_a<H_{c1}$; see Sec.~\ref{IIIB}. However, under condition (\ref{5c}), the sheet current formally diverges  in  the immediate vicinity of the left corners, and the shrinking segments can get stuck  there. Then, small domes of the inclined vortices should appear near the left corners in order to suppress the divergence of the current and, thereby, allow  the vortices to leave the sample. We have ignored these inclined vortices in our analysis since they hardly change condition (\ref{5c}) and have a little effect on the currents flowing at $u>u_2^{(0)}$ (the dome of the inclined vortices is small as compared to the main vortex dome on the upper surface of the strip). The existence of the inclined vortices also does not disturb the vortex-entry condition imposed on the currents on the opposite lateral surface [the current density near the right corners, Eq.~(\ref{27}), remains practically unchanged at a small variation of $u_1^{(0)}$ if $u_1^{(0)}\approx -1$].
However, the inclined vortices can modify the current $I_L$ flowing on the left lateral surface, and at $H_a\sim H_{c1}$ [when $|I_L|\sim I$ according to Eq.~(\ref{25})], they generally have to be taken into account in the calculation of the critical current.

A similar estimates can be carried out in the case when the curved vortex at the right boundary of the dome touches the right equatorial point. This touch occurs at $H_a=H_{\rm tch}\sim H_{c1}$, and therefore one may expect that in the range $H_a>H_{\rm tch}$, the inclined vortices exist on the right lateral surface of the strip since there is no free room in the bulk of the strip for the penetrating vortices. These inclined vortices are likely to affect the vortex-entry condition imposed on the currents flowing on the same lateral surface [$j_{\rm crn}$ near the right corners, Eq.~(\ref{27}), is sensitive to $u_2^{(0)}$ if $u_2^{(0)}\to 1$], and in fact, a new scenario of the vortex penetration occurs in this magnetic-field range. Therefore, the $H_a$-dependence of the critical current at $H_a>H_{\rm tch}$  requires an additional study. It is also worth noting that for $p/p_c<1$, the crossing point $H_{cr}$ of the functions $I_c^{\rm GB}(H_a)$ and $I_c^{\rm BL}(H_a)$ (Fig.~\ref{fig11}) approaches $H_{c1}$ with decreasing the ratio $p/p_c$, and $H_{cr}$ can exceed the field at which the right boundary of the dome touches the right equatorial point. For such values of $p$, the geometrical barrier prevails over the Bean-Livingston one up to $H_{\rm tch}$, and the crossover in $I_c(H_a)$ discussed in Sec.~\ref{IVc} does not occur.

\section{Conformal mapping and sheet currents in the Meissner state of a strip placed in a magnetic field}\label{SM}

\begin{figure}[tbh] 
 \centering  \vspace{+9 pt}
\includegraphics[scale=.41]{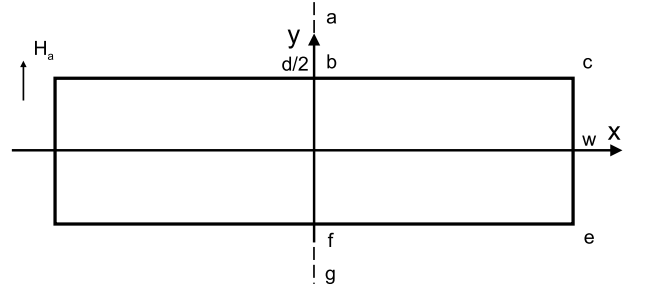}
\caption{\label{fig1s} The magnetic-field line ($a$-$b$-$c$-$e$-$f$-$g$) adjoining a superconducting strip in the Meissner state. Under the conformal mapping described in the text, this line is the image of the real axis, $A=0$, of the complex plane $\varphi-iA$.
 } \end{figure}

For an infinitely long strip in the Meissner state, the magnetic field ${\bf H}(x,y)$ outside the sample can be found from the Maxwell equations
 \[
{\rm div}{\bf H}=0, \ \ \ \ {\rm rot}{\bf H}=0,
 \]
and hence the field can be described both by the scalar potential
$\varphi(x,y)$, ${\bf H}=-\nabla \varphi$, and by the vector
potential ${\bf A}={\bf z} A(x,y)$, ${\bf H}={\rm rot}{\bf A}$,
where ${\bf z}$ is the unit vector along the $z$ axis (i.e., along the axis of the strip). The complex potential $F=\varphi-i A$ is an analytical function of $x+iy$ (and $x+iy$ is an analytical function of $\varphi-i A$) \cite{LL}, and it can be found with the appropriate conformal mapping. Knowing the potential, one can calculate the surface Meissner current in the strip. As an example, consider the strip with a rectangular cross section and find the sheet currents flowing near its surface in the case when the strip is placed in the external magnetic field $H_a$ (Fig.~\ref{fig1s}). Let the width and the thickness of the strip be $2w$ and $d$, respectively.

\subsection{Conformal mapping}

To calculate the complex potential, let us find the conformal mapping of the upper half of the complex plane $F=\varphi-i A$ to the region lying to the right of line $a$-$b$-$c$-$e$-$f$-$g$ in Fig.~\ref{fig1} (the contour $a$-$b$-$c$-$e$-$f$-$g$ coincides with the magnetic-field line $A=0$ in the $x-y$ plane). This region is the rectangle with one of its sides, $f$-$g$-$a$-$b$, passing through the infinite point. The angles $\alpha_i\pi$ of the rectangle  at the vertices $b$ and $f$ are equal to $\pi/2$, i.e.,  $\alpha_b=\alpha_f=1/2$, whereas  at the vertices $c$ and $e$, they are $3\pi/2$, and so $\alpha_c= \alpha_e=3/2$. The mapping can be found using the Schwarz-Christoffel formula \cite{tfkp},
\begin{eqnarray} \label{1s}
&x&+iy=C_0+\\
\!&&\!\!C\!\!\int_0^{F}\!\!\!\!(t-t_b)^{\alpha_b-1}\!(t-t_c)^{\alpha_c-1}\!
(t-t_e)^{\alpha_e-1}\!(t-t_f)^{\alpha_f-1}dt, \nonumber
\end{eqnarray}
where $t$ is the complex variable in the plane $\varphi-i A$; $x+iy$ is the image of a point $F$ lying in the upper half of this plane; $t_i$ are points in the real axis $A=0$ which map onto the vertices $b$, $c$, $e$, $f$  of the rectangle; $C$ and $C_0$ are some constants. It is convenient to choose $t_c=-1$, $t_e=1$, $t_b=-1/k$, $t_f=1/k$ where $k=\sqrt{m}<1$, and $m$ is the parameter which will be found below. In other words, we consider $t$ as the dimensionless variable of  the $\varphi-iA$ plane, $t\equiv (\varphi-iA)/\varphi_e$, where the constant $\varphi_e$ is the scalar potential at the point $e$.  With these $t_i$ and $\alpha_i$, formula (\ref{1s}) is rewritten as follows:
\begin{eqnarray} \label{2s}
x+iy=C_1\int_0^{F}\frac{(1-t^2)^{1/2}}{(1-k^2t^2)^{1/2}}dt +C_0,
\end{eqnarray}
where $C_1=Ck$. Now let us find the constants $C_1$, $C_0$, and the parameter $m$.

It is clear from the symmetry that the equatorial point ($x=w$,$y=0$) of the strip corresponds to the point $t=0$ in the axis $A=0$. Then, setting $F=0$ in Eq.~(\ref{2s}), we obtain,
\begin{eqnarray} \label{3s}
w+i0=C_0,
\end{eqnarray}
i.e., $C_0=w$. On the other hand, the point ($x=w$,$y=-d/2$) corresponds to the point $t=1$ in the axis $A=0$. Therefore,
\begin{eqnarray} \label{4s}
w-i\frac{d}{2}=C_1\int_0^{1}\frac{(1-t^2)^{1/2}}{(1-k^2t^2)^{1/2}}dt +w,
\end{eqnarray}
and so
\begin{eqnarray} \label{5s}
C_1=-i\frac{d}{2}\left[\int_0^{1}\frac{(1-t^2)^{1/2}}{(1-k^2t^2)^{1/2}}dt \right]^{-1}=-i\frac{d}{2}\frac{m}{f(1,m)},~~~~
\end{eqnarray}
where we have introduced the notation,
\begin{eqnarray} \label{6s}
f(u,m)\equiv m\int_0^{u}\frac{(1-t^2)^{1/2}}{(1-mt^2)^{1/2}}dt.
\end{eqnarray}
Taking into account that the point ($x=0$,$y=-d/2$) corresponds to the point $t=1/k$ in the axis $A=0$, we obtain the relation defining the parameter $m$,
\begin{eqnarray} \label{7s}
-i\frac{d}{2}&=&C_1\int_0^{1}\frac{(1-t^2)^{1/2}}{(1-k^2t^2)^{1/2}}dt \nonumber \\&+&C_1\int_1^{1/k}\frac{(1-t^2)^{1/2}}{(1-k^2t^2)^{1/2}}dt+w.
\end{eqnarray}
With Eq.~(\ref{5s}), this relation yields formula (4) presented in the main text,
\begin{eqnarray} \label{8s}
\frac{2w}{d}=\frac{m}{f(1,m)}\int_1^{1/k}\!\!\!
\frac{(t^2-1)^{1/2}}{(1-k^2t^2)^{1/2}}dt= \frac{f(1,1-m)}{f(1,m)},~~~
\end{eqnarray}
where the integral over $t$ from $1$ to $1/k$ has been transformed  with the following change of the variable: $t=\sqrt{1-(1-k^2)v^2}/k$.

For the points ($x=w$,$-d/2\le y\le d/2$) on the right lateral surface  of the strip ($-1\le t \le 1$), formula (\ref{2s}) gives,
\begin{eqnarray*}
w+iy=C_1\int_0^{t}\frac{(1-v^2)^{1/2}}{(1-k^2v^2)^{1/2}}dv +w,
\end{eqnarray*}
and hence we arrive at,
\begin{eqnarray}\label{9s}
\frac{2y}{d}=-\frac{m}{f(1,m)}\int_0^{t}\frac{(1-v^2)^{1/2}}{(1-k^2v^2)^{1/2}}dv =\frac{f(-t,m)}{f(1,m)}.~~~
\end{eqnarray}
In the main text, the variable $u$ is introduced that parameterizes the points ($0\le x \le w$,$y=d/2$) on the upper surface of the strip ($0\le u\le 1$) and the points ($x=w$,$0\le y\le d/2$) on the right lateral surface ($1\le u\le 1/\sqrt{1-m}$). This $u$ is related to  the variable $t$ as follows:
 \begin{eqnarray}\label{10s}
  t=-\frac{\sqrt{1-(1-m)u^2}}{\sqrt{m}}.
 \end{eqnarray}
Formulas (\ref{9s}), (\ref{10s}) reproduce expressions (6), (7) for the function $y(u)$ presented in the main text.

For the points ($0\le x\le w$,$y=d/2$) on the upper surface of the strip ($-1/k\le t \le -1$), formula (\ref{2s}) gives,
\begin{eqnarray*}
x+i\frac{d}{2}&=&C_1\int_0^{-1}\frac{(1-v^2)^{1/2}}{(1-k^2v^2)^{1/2}}dv \\ &+&C_1\int_{-1}^{t}\frac{(1-v^2)^{1/2}}{(1-k^2v^2)^{1/2}}dv+w,
\end{eqnarray*}
and hence we arrive at,
\begin{eqnarray}\label{11s}
&&\frac{x}{w}=-\frac{d}{2w}\frac{m}{f(1,m)}\int_{1}^{-t} \frac{(v^2-1)^{1/2}}{(1-k^2v^2)^{1/2}}dv +1 \nonumber \\ &=&\frac{d}{2w}\frac{m}{f(1,m)}\int_{-t}^{1/k}\!\!\! \frac{(v^2-1)^{1/2}}{(1-k^2v^2)^{1/2}}dv=\frac{f(u,1-m)}{f(1,1-m)},~~~~~~~
\end{eqnarray}
where we have used both the equalities in Eq.~(\ref{8s}), relation (\ref{10s}), and the integral over $v$ from $-t$ to $1/k$ has been transformed with the following change of the variable, $v= \sqrt{1-(1-k^2)v_1^2}/k$. Formula (\ref{11s}) reproduces expression (2) for the function $x(u)$ given in the main text.

To find the constant $\varphi_e$, let us set $F=t\gg 1,1/k$ in Eq.~(\ref{2s}), i.e., consider a point ($x=0$,$y$) that lies in the axis $x=0$ and is far away from the strip ($|y|\gg d, w$). Then, the differentiation of Eq.~(\ref{2s}) over $y$ gives
\begin{eqnarray*}
i=C_1\frac{(1-t^2)^{1/2}}{(1-k^2t^2)^{1/2}}\frac{dt}{dy}\to \frac{C_1}{k \varphi_e} \frac{d\varphi}{dy} \to -\frac{C_1}{k \varphi_e}H_a,
\end{eqnarray*}
where we have taken into account that at $t=(\varphi/\varphi_e)\to \infty$, the magnetic field $-{d\varphi}/{dy}$ tends to $H_a$. Hence,
\begin{eqnarray}\label{12s}
\varphi_e=-\frac{C_1}{i k}H_a,
\end{eqnarray}

\subsection{Surface currents}

Calculating ${\bf H}=-\nabla \varphi$ with the use of the obtained potential at the surface of the strip (${\bf H}$ is tangential to the surface in the Meissner state), one can finds the Meissner sheet currents $J_M=J_z$ flowing near this surface in the layer of the thickness  $\sim \lambda$,
\begin{eqnarray} \label{13s}
{\bf J}_M=[{\bf n}\times {\bf H}],
\end{eqnarray}
where ${\bf n}$ is the outward normal to the surface of the sample  at the point of interest \cite{LL}.
To find the currents on the right lateral surface of the strip, we set $F=t$ in formula (\ref{2s}) where $|t|\le 1$, differentiate this formula over $y$, and use relation (\ref{12s}),
\begin{eqnarray*}
i&=&C_1\frac{(1-t^2)^{1/2}}{(1-k^2t^2)^{1/2}}\frac{dt}{dy}= C_1\frac{(1-t^2)^{1/2}}{(1-k^2t^2)^{1/2}}\frac{(-H_y)}{\varphi_e} \\
&=&ik\frac{(1-t^2)^{1/2}}{(1-k^2t^2)^{1/2}}\frac{J_M}{H_a}.
\end{eqnarray*}
Thus, this formula gives,
\begin{eqnarray} \label{14s}
\frac{ J_M(t)}{H_a}=\frac{\sqrt{1-mt^2}}{\sqrt
m\sqrt{1-t^2}}\,.
\end{eqnarray}
In a similar manner, one can find the currents on the upper surface of the strip, differentiating formula (\ref{2s}) over $x$. Eventually, we again arrive at Eq.~(\ref{14s}), with $1-t^2$ being replaced by $t^2-1$ (now $-1/k\le t \le -1$). Using Eq.~(\ref{10s}), formula (\ref{14s}) can be rewritten in terms of the variable $u$, 
 \begin{eqnarray} \label{15s}
\frac{ J_M(t)}{H_a}=\frac{u}{\sqrt{|1-u^2|}}\,.
 \end{eqnarray}
As a result, we obtain the expression which is applicable for $|u|\le 1/\sqrt{1-m}$ and reproduces Eq.~(9) in the main text in the case when the strip is placed in the external magnetic field $H_a$.

{}


\begin{thebibliography}{}
\bibitem{com1} For $u$ lying in the interval $0\le u \le 1/\sqrt{1-m}$, the parameter $t$ of Ref.~\cite{jetp13} is expressed in terms of $u$ as follows: $t=-\sqrt{1-(1-m)u^2} /\sqrt{m}$, and one has $1/\sqrt{m}\ge |t| \ge 0$.
\end{thebibliography}
\end{document}